\documentclass[12pt]{article}
\usepackage{pdfpages}
\usepackage{wrapfig}
\usepackage{bm}
\usepackage{upgreek}
\usepackage{dsfont}
\usepackage{simplewick}
\usepackage{mathtools}
\usepackage{framed}
\usepackage{microtype}
\usepackage{marvosym}
\usepackage{color}
\usepackage{transparent}
\usepackage{placeins}
\usepackage{mdframed,mdwlist}
\usepackage{graphicx}
\usepackage{float}
\usepackage{longtable}
\usepackage{mathdots}
\usepackage{eucal}
\usepackage{array}
\usepackage{stmaryrd}
\usepackage{amsthm}
\usepackage{pifont}
\usepackage{lipsum}
\usepackage{enumerate}
\usepackage{amsmath}
\usepackage{amssymb}
\usepackage{algorithm}
\usepackage{algpseudocode}
\usepackage{setspace}
\usepackage{gensymb}
\usepackage{adjustbox}
\usepackage{orcidlink}
\usepackage[authoryear]{natbib}
\usepackage{orcidlink}
\DeclareMathOperator*{\argmin}{arg\,min}
\DeclareMathOperator*{\plim}{plim\,}

\newtheorem{theorem}{Theorem}
\newtheorem{proposition}{Proposition}
\newtheorem{lemma}{Lemma}
\newtheorem{corollary}{Corollary}

\begin{document}
\title{Consistent information criteria for regularized regression and loss-based learning problems}
\author{Qingyuan Zhang$^{1\,\orcidlink{0000-0002-0906-8533}}$, Hien Duy Nguyen$^{2,3\,\orcidlink{0000-0002-9958-432X}}$}
\date{}

\maketitle

\noindent\footnotesize{$^{1}$School of Mathematics and Physics, University of Queensland, St Lucia, QLD 4072, Australia\\
$^{2}$School of Computing, Engineering and Mathematical Sciences, La Trobe University, Bundoora, VIC 3086, Australia\\
$^{3}$Institute of Mathematics for Industry, Kyushu University, Nishi Ward, Fukuoka 819-0395, Japan\\
Email: h.nguyen5@latrobe.edu.au}

\doublespacing
\section*{Abstract}
Many problems in statistics and machine learning can be formulated as model selection problems, where the goal is to choose an optimal parsimonious model among a set of candidate models. It is typical to conduct model selection by penalizing the objective function via information criteria (IC), as with the pioneering work by Akaike and Schwarz. Via recent work, we propose a generalized IC framework to consistently estimate general loss-based learning problems. In this work, we propose a consistent estimation method for Generalized Linear Model (GLM) regressions by utilizing the recent IC developments. We advance the generalized IC framework by proposing model selection problems, where the model set consists of a potentially uncountable set of models. In addition to theoretical expositions, our proposal introduces a computational procedure for the implementation of our methods in the finite sample setting, which we demonstrate via an extensive simulation study.

\noindent \textbf{Keywords}: Information criteria; model selection; generalized linear models; least absolute shrinkage and selection operator; regularization; asymptotic theory

\section{Introduction}
Model selection is a common problem in many statistical learning and machine learning applications. The typical objective is to select an appropriate model from a set of candidate models on the basis of sample data. Model selection criteria provide a methodology for achieving such a goal, where a good criterion defines a trade-off between the quality of a model's fit to the sample data and the complexity of the model. Archetypal examples of such criteria are the Akaike Information Criteria (AIC; \citealt{akaike_new_1974}) and the Bayesian Information Criteria (BIC; \citealt{schwarz_estimating_1978}), which are among the earliest proposed methods and remain among the most popular approaches. Detailed expositions on the subject of model selection can be found in the volumes of \cite{mcquarrie_regression_1998}, \cite{ vapnik_statistical_1998}, \cite{vapnik_nature_2000}, \cite{claeskens_model_2001}, and \cite{buhlmann_statistics_2011}. 

Our study focuses on establishing model selection schemes that provide consistency guarantees. The seminal work of \cite{sin_information_1996} provides consistency results for a broad class of IC for generic loss-based learning problems. However, their theorems impose strong conditions on the objective function and the candidate models that are challenging to verify in practice. In \cite{nguyen_panic:_2023}, the IC of \cite{sin_information_1996} for generic risk minimization problems are further studied, yielding a generic framework for model selection, which we refer to as PanIC. The PanIC framework, together with the works of \cite{sin_information_1996} and \cite{baudry2015estimation}, consider IC in generic loss-based learning problems. This contrasts with other works in the literature, which typically assume the learning objective to be a negative log-likelihood function \citep{leroux_consistent_1992, hui_order_2015}, or a negative composite, pseudo, or quasi log-likelihood function \citep{varin_note_2005, gao_composite_2010, ng_model_2014, hui_use_2021}. 

In this work, we seek to specialize and refine the PanIC framework towards consistent estimation methods for general regression problems. Consistency conditions for parameter estimators and minimal risks of regression problems under the Empirical Risk Minimization (ERM) approach are well established and can be found, for example, in \cite{vapnik_statistical_1998},  \cite{vapnik_nature_2000}, and \cite{shapiro_lectures_2021}. However, it is observed that
learning model complexity according to the ERM principle tends to produce models that overfit the data. This fact has been widely recognized in the literature and has motivated the broad study of regularization techniques. Examples of such methods include $l_2$ regularization, leading to the Ridge regression problem \citep{hoerl_ridge_1970}; $l_1$ regularization, leading to the Lasso problem \citep{tibshirani_regression_1996}; and a convex combination of of $l_2$ and $l_1$ penalties, leading to the elastic-net problem \citep{zou_regularization_2005}. Results regarding the consistency of estimators obtained from solving such regularization problems, and others, can be found, for example, in \cite{zhao_model_2006}, \cite{yuan_non-negative_2007}, \cite{jia_model_2010}, and \cite{hastie_statistical_2015}.

To address the consistency of regularized regression problems, we extend the PanIC framework from the finite set of models setting of \cite{nguyen_panic:_2023}, to be able to handle sets of models that may be uncountably large. As we will argue in Section \ref{Sect: Method}, in many learning problems, considering infinitely many models is both natural and advantageous. It is therefore useful to derive conditions under which consistency may be proved in such settings. 

A summary of our contributions is as follows:
\begin{enumerate}
    \item We outline sufficient conditions for consistently estimating various regression problems, including linear, logistic, Poisson, and Gamma regressions. Notably, we provide details of our derivations that can be followed by the reader to derive conditions for regression problems not addressed in this work.
    \item We expand upon the theory of PanIC by demonstrating its consistency in certain model selection problems involving infinitely many models. Although not expansive enough to cover all model selection problems on uncountable 
    spaces of models, we argue that our formulation is nevertheless natural for a ubiquitous class of learning problems and is thus of broad practical interest. 
    \item We propose a computational method for implementing PanIC in regression problems, and analyze the performance of our routine against popular benchmarks in finite sample studies. We note that our method maintains the asymptotic consistency as a PanIC estimator and performs effectively in simulated regression problems.
\end{enumerate}
We note in particular that our work differs from previous works in that the consistency condition we derive concerns the norm of the optimal model, rather than consistency in terms of parameter estimation or model selection, as per \cite{fan_2001} and \cite{zhao_model_2006}. A related problem appears in the literature in the work of \cite{massart_2011_lasso_l1}. However, the emphasis of \cite{massart_2011_lasso_l1} is on the oracle inequalities related to the Lasso problem, while our focus lies on the asymptotic consistency of a more general set of regularized regression problems without first obtaining finite sample oracles.

The rest of this article proceeds as follows: in Section \ref{Sect: Method}, the PanIC framework is reviewed and our proposed estimation method is given with theoretical justification. In Section \ref{Sect: Simulation}, a simulated study of PanIC in linear regression problems is reported with some discussions. Finally, concluding remarks and an outline of future research directions are given in Section \ref{Sect: Conclusion}.

\section{Method} \label{Sect: Method}
We review the PanIC framework in Section \ref{Sect: PanIC}. The connection between regularized regression estimation and PanIC is established in Section \ref{Sect: Regression}. A formal justification of our proposed estimation method is presented in Section \ref{Sect: Theory}. Lastly, a computational method is given in Section \ref{Sect: Computations}. 

\subsection{PanIC}\label{Sect: PanIC}
Consider the following setting of a general model selection problem. Let $(\Omega, \mathcal{F}, \mathbb{P})$ be the underlying probability space and $X: \Omega \to \mathbb{X} \subset \mathbb{R}^d$ be a random vector on the pushforward probability space $(\mathbb{X}, \mathcal{B}(\mathbb{X}), \mathbb{P}_X)$, where $\mathcal{B}(\mathbb{X})$ denotes the Borel $\sigma$-algebra of $\mathbb{X}$. We observe a sequence of independent and identically distributed (i.i.d) random variables $(X_i)_{i \in \left[n\right]}$ on the probability space $(\mathbb{X}, \mathcal{B}(\mathbb{X}), \mathbb{P}_{X})$, where $\left[n\right] = \{1, \hdots, n\}$. The set of candidate models is given by a sequence of hypotheses $(\mathcal{H}_k)_{k \in \left[m\right]}$. Each $\mathcal{H}_k$ defines a functional space identified by some parameter vector $\beta_k \in \mathbb{T}_k \subset \mathbb{R}^{q_k}$, where $(q_k)_{k \in \left[m\right]} \subset \mathbb{N}$. That is, for each $k \in \left[m\right]$, 
\[\mathcal{H}_k = \{h_k(\cdot; \beta_k): \mathbb{X} \to \mathbb{R}: \beta_k \in \mathbb{T}_k\}.\]

This construction admits two sources of flexibility. For each hypothesis $\mathcal{H}_k$ with index $k \in \left[m\right]$, the functional form of $h_k$ and the parameter space $\mathbb{T}_k$ are both allowed to vary. We further define a loss function $\ell: \mathbb{R} \to \mathbb{R}$, and for each $k \in \left[m\right]$, denote, 
\[R_{k,n}(\beta_k) = \frac{1}{n} \sum^n_{i=1}\ell(h_k(X_i; \beta_k))\]
and 
\[r_k(\beta_k) = \mathbb{E} \{\ell(h_k(X; \beta_k))\}\]
as the empirical risk and the expected risk, respectively. We shall abbreviate $\ell(h_k(X; \beta_k))$ by $\ell_k(X; \beta_k)$, which implicitly defines $\ell_k(\cdot; \beta_k): \mathbb{X} \to \mathbb{R}$ and $\ell_k(x; \cdot): \mathbb{T}_k \to \mathbb{R}$ for each hypothesis $\mathcal{H}_k$ with index $k \in \left[m\right]$. Examples of some classic model selection problems under this formulation are presented by \cite{nguyen_panic:_2023}.

Further, we call $k^*$ the optimal hypothesis index,
\begin{equation*}\label{Eqs: Optimal hypothesis index}
    k^* = \min\argmin_{k \in \left[m\right]} \left\{\min_{\beta_k \in \mathbb{T}_k}r_k(\beta_k)\right\}.
\end{equation*}
The corresponding hypothesis space $\mathcal{H}_{k^*}$ is known as the class of parsimonious models. Let $\hat{K}_n$ denote an estimator of $k^*$. A model selection scheme is said to be consistent if 
\begin{equation} \label{Eqs: Consistency}
    \plim_{n\to\infty} \hat{K}_n = k^*,
\end{equation}
where $\plim$ denotes convergence in probability (as per \citealp[Sec. 3.2]{amemiya1985advanced}). In the case where the set of candidate models is finite, this is equivalent to 
\begin{equation*}\label{Eqs: Probability Limit}
    \lim_{n\to\infty}\mathbb{P}\left(\hat{K}_n = k^*\right) = 1.
\end{equation*}

Now, suppose that the model selection problem satisfies conditions A1--A3, for each $k \in \left[m\right]$:
\begin{enumerate}
    \item[A1]: $\ell_k(x; \beta_k)$ is Caratheodory in the sense that $\ell_k(x; \cdot): \mathbb{T}_k \to \mathbb{R}$ is continuous for each $x\in \mathbb{X}$, and $\ell_k(\cdot; \beta_k): \mathbb{X} \to \mathbb{R}$ is $\mathcal{B}(\mathbb{X})$-measurable for each $\beta_k\in\mathbb{T}_k$.
    \item[A2]: $\mathbb{T}_k$ is compact and there exists some $\tau_k\in\mathbb{T}_k$ such that $\ell_k(X; \tau_k)^2$ is square integrable. That is,
    \[\mathbb{E}\{\ell_k(X;\tau_k)^2\} < \infty.\]
    \item[A3]: There exists some measurable function $\mathcal{G}_k:\mathbb{X}\to \mathbb{R}_{\geq 0}$, such that $\mathbb{E}\{\mathcal{G}_k(X)^2\} < \infty$ and, 
    \[|\ell_k(x; \beta_k) - \ell_k(x;\tau_k)| \leq \mathcal{G}_k(x) \lVert \beta_k - \tau_k\rVert, \quad (\forall \beta_k, \tau_k\in\mathbb{T}_k, \text{ a.e. } x\in\mathbb{X}).\]
\end{enumerate}

A PanIC estimator $\hat{K}_n$ is defined as
\begin{equation}\label{Eqs: PanIC estimator}
    \hat{K}_n = \min \argmin_{k \in \left[m\right]} \left\{\min_{\beta_k \in \mathbb{T}_k}R_{k,n}(\beta_k) + P_{k,n}\right\},
\end{equation}
where $P_{k,n}: \Omega \to \mathbb{R}_{\geq 0}$ is allowed to be a stochastic, non-negative function that satisfies conditions B1 and B2, for each $k \in \left[m\right]$:
\begin{enumerate}
    \item[B1]: $P_{k, n} > 0$, for $n \in \mathbb{N}$, and $P_{k, n}= o_{\mathbb{P}}(1)$ as $n\to\infty$.
    \item[B2]: If $k < l$, then $\sqrt{n}\{P_{l, n} - P_{k, n}\} \xlongrightarrow{\mathbb{P}} \infty$, as $n\to\infty$.
\end{enumerate}
Under these assumptions, Theorem 1 of \cite{nguyen_panic:_2023} shows that PanIC estimators are consistent. Equivalently, an estimator taking the form of (\ref{Eqs: PanIC estimator}) consistently solves the model selection problem, provided that conditions A1--A3, B1, and B2 are satisfied. 

\cite{nguyen_panic:_2023} also introduced the notion of a BIC-like criterion, one that satisfies the condition $\text{B2}^*$ instead of B2,
\begin{enumerate}
    \item[$\text{B2}^*$]: If $k < l$, then $n\{P_{l, n} - P_{k, n}\} \xlongrightarrow{\mathbb{P}} \infty$, as $n\to\infty$,
\end{enumerate}
Theorem 2 of \cite{nguyen_panic:_2023} shows the consistency of BIC-like criteria under additional assumptions C1--C5. For $k \in \left[m\right]$, 
    \begin{enumerate}
        \item [C1]: $r_k$ is Lipschitz continuous on $\mathbb{T}_k$, and it is twice differentiable and uniquely minimized at some $\beta^*_k\in\mathbb{T}_k$. 
        \item [C2]: $\ell_k(x, y;)$ is Lipschitz continuous and differentiable at $\beta^*_k$, for almost all $(x, y)\in\mathbb{X}$.
        \item [C3]: The set $\mathbb{T}_k$ is second order regular at $\beta^*_k$.
        \item [C4]: The quadratic growth condition holds at $\beta^*_k$.
        \item [C5]: If $r_k(\beta^*_k) = r_{k^*}(\beta^*_{k^*})$, then $n(R_{k,n}(\beta^*_k) - R_{k,n}(\beta^*_{k^*})) = O_{\mathbb{{P}}}(1)$.
    \end{enumerate}

As an important point to be revisited in Section \ref{Sect: Simulation}, we note that conditions B1, B2, and $\text{B2}^*$ remain satisfied when the penalty function $P_{k,n}$ is multiplied by any positive constant. We will express a valid PanIC penalty in the form
\begin{equation}\label{Eqs: Penalty Shape}
    P_{k, n} = \kappa \times \text{pen}_{\text{shape}}(k,n)
\end{equation}
and call $\text{pen}_{\text{shape}}:\mathbb{N}\times\mathbb{N}\rightarrow \mathbb{R}_+$ the penalty shape of $P_{k,n}$. From this expression, it is clear that any valid PanIC penalty is defined up to a multiplicative constant. This observation holds practical significance, as it implies that the constant $\kappa$ should be treated as a hyperparameter, requiring calibration in finite sample settings. 

\subsection{Regression Problems} \label{Sect: Regression}
This section connects the regression estimation problem with a model selection problem in the PanIC framework. In regression problems, we observe some response $y\in \mathcal{Y}\subset \mathbb{R}$, and covariate vector $x\in\mathcal{X}\subset \mathbb{R}^d$, for some positive integer $d$. Each $x\in \mathcal{X}$ is assumed to be related to a $y\in\mathcal{Y}$ through some stochastic dependencies. The goal is to infer from an observed sample $(x_i, y_i)_{i\in\left[n\right]}$ this underlying relationship. In the Generalized Linear Model (GLM) framework, this relationship is assumed to be defined by a single functional form $h(\cdot; \beta_0, \beta): \mathcal{X} \to \mathcal{Y}$ that is parametrized by a coefficient vector $\beta$ taking values in $\mathbb{R}^d$ and a bias term $\beta_0$ taking values in $\mathbb{R}$. Further, we assume the existence of a loss function $\ell: \mathcal{X} \times \mathcal{Y} \to \mathbb{R}$. Our focus is on linear, logistic, Poisson, and Gamma regressions. These regressions collectively model a comprehensive range of common response types, including continuous, binary, count, and positive continuous data. Table \ref{Table: Regression} reviews these regression problems,
\begin{table}[H]
\caption{Mean and loss functions for generalized linear regression models. Here, $h=h(x;\beta_0,\beta)$.}
\label{Table: Regression}
\centering
\footnotesize
\begin{tabular}{c c c} 
 Regression & Mean function & Loss function\\
  [0.5ex] 
  \hline\hline
 Linear & $h = \beta_0 + \beta^\top x$ & $\ell(x, y; \beta_0, \beta) = \left(y - h\right)^2$ \rule{0pt}{5ex}\\
 Logistic & $h = (1 + \exp(-(\beta_0 + \beta^\top x)))^{-1}$ & $ \ell(x, y; \beta_0, \beta) = -\left\{y\log h
     + (1 - y)\log\left(1 - h\right)\right\}
$ \rule{0pt}{5ex}\\
 Poisson & $h = \exp(\beta_0 + \beta^\top x)$ & $
 \ell(x, y; \beta_0, \beta) = -\left\{y\log h - h
 -\log(y!) \right\} $ \rule{0pt}{5ex}\\
 Gamma & $h = \exp(\beta_0 + \beta^\top x)$ & $
 \ell(x, y; \beta_0, \beta, \nu) = -\left\{ -\nu\log(h)
  + (\nu - 1) \log (y)
 - y\nu/h\right\} $ \rule{0pt}{5ex}\\[4ex]
 \hline\hline
\end{tabular}
\end{table}
\noindent Following the GLM convention, the loss function of each regression is defined by the negative log-likelihood of the corresponding distribution. Specifically, for the Gamma regression, we use the mean parameterization of the Gamma density function
\[p(y;\mu, \nu) = \frac{1}{\Gamma(\nu)}\left(\frac{\nu}{\mu}\right)^\nu y^{\nu-1}\exp\left(-y\frac{\nu}{\mu}\right)\text{,}\]
where $\mathbb{E}(Y) = \mu$ and $\text{Var}(Y) = \mu^2/\nu$.

Under the Empirical Risk Minimization principle (ERM), a regression problem is estimated by minimizing the empirical risk over the parameter space. Assuming a unique solution exists, learning according to the ERM principle yields the estimates $\hat{\beta}_0$ and $\hat{\beta}$, where
\[
\left(\hat{\beta}_{0},\hat{\beta}\right)=\underset{\beta_{0}\in\mathbb{R},\beta\in\mathbb{R}^{d}}{\arg\min}~\frac{1}{n}\sum_{i=1}^{n}\ell\left(x_{i},y_{i};\beta_{0},\beta\right)\text{.}
\]
To correct the overfitting tendency of the ERM, the Lasso regression, the Ridge regression, or the Elastic-net regression are often solved instead. We note that both Lasso and Ridge regression problems can be represented in a unified manner with the following formulation. Define 
\begin{align}\label{Eqs: Lasso Ridge}
\begin{split}
    \mathcal{H} &= \{h(\cdot; \beta_0, \beta): \mathcal{X} \to \mathcal{Y}, (\beta_0, \beta) \in \mathbb{T}\}\text{,}\\
    \mathbb{T} &= \{(\beta_0, \beta) \, |\, \beta_0 \in \mathbb{U},\, \beta\in \mathbb{R}^d: \lVert\beta\rVert_p \leq C, \, C \in \mathbb{R}_+\},
\end{split}
\end{align}
where $\mathbb{U}$ is some sufficiently large compact subset of $\mathbb{R}$ that contains the optimal bias term $\beta^*_0$ in its interior. Setting $p=1$ in (\ref{Eqs: Lasso Ridge}) corresponds to the Lasso problem \citep{tibshirani_regression_1996}, and setting $p=2$ corresponds to the Ridge regression problem \citep{hoerl_ridge_1970}. 
Slightly modifying (\ref{Eqs: Lasso Ridge}), we also retrieve the Elastic-net problem
\citep{zou_regularization_2005}:
\begin{align}\label{Eqs: Elastic Net}
\begin{split}
    \mathcal{H} &= \{h(\cdot; \beta_0, \beta): \mathcal{X} \to \mathcal{Y}, (\beta_0, \beta) \in \mathbb{T}\}\text{,}\\
    \mathbb{T} &= \{(\beta_0, \beta) \, |\, \beta_0 \in \mathbb{U},\, \beta\in \mathbb{R}^d: \alpha\lVert\beta\rVert_1 + (1-\alpha)\lVert\beta\rVert_2^2 \leq C, \, C \in \mathbb{R}_+\}, 
\end{split}
\end{align}
for some $\alpha \in (0, 1)$. Notably, both (\ref{Eqs: Lasso Ridge}) and (\ref{Eqs: Elastic Net}) are equivalent to a single hypothesis within the PanIC framework, dependent on a hyperparameter $C$. If we consider a sequence of positive real numbers denoted by $(C_k)_{k\in\left[m\right]}$, we obtain a sequence of hypotheses:
\begin{align}\label{Eqs: Model selection}
\begin{split}
    \mathcal{H}_k &= \{h(\cdot; \beta_0, \beta): \mathcal{X} \to \mathcal{Y}, (\beta_0, \beta) \in \mathbb{T}_k\}\\
    \mathbb{T}_k &= \begin{cases}
        \{(\beta_0, \beta) \, |\, \beta_0 \in \mathbb{U}, \beta\in \mathbb{R}^d: \lVert\beta\rVert_1 \leq C_k\}, & \text{for Lasso regression,}\\
        \{(\beta_0, \beta) \, |\, \beta_0 \in \mathbb{U}, \beta\in \mathbb{R}^d: \lVert\beta\rVert_2 \leq C_k\}, & \text{for Ridge regression,}\\
        \{(\beta_0, \beta) \, |\, \beta_0 \in \mathbb{U}, \beta\in \mathbb{R}^d: \alpha\lVert\beta\rVert_1 + (1-\alpha)\lVert\beta\rVert_2^2 \leq C_k\}, & \text{for Elastic-net regression}.
    \end{cases}
\end{split}
\end{align}
Then, utilizing the theory of PanIC, we can construct a consistent estimator of the optimal hypothesis index $k^*$. Thus, the PanIC framework provides a unifying approach to consistently solve constrained regression problems.

\subsection{Consistency} \label{Sect: Theory}
We outline sufficient conditions for consistent model selection in linear, logistic, Poisson, and Gamma regression problems. These conditions are derived from fulfilling conditions A1--A3 of PanIC, with complete derivations given in the Appendix.
\begin{proposition}\label{Prop: Sufficient conditions}
Assume that $(X_i, Y_i)_{i\in\left[n\right]}$ is an i.i.d sequence and specific conditions are met for each regression problem. Then, the PanIC estimator satisfies the consistency property (Eq. \ref{Eqs: Consistency}) in the corresponding problem given the specific assumptions.
    \begin{enumerate}
            \item [1.] Linear regression: Both the covariate vector $X$ and the response $Y$ have a finite fourth moment.
            \item [2.] Logistic regression: The covariate vector $X$ has a finite second moment. 
            \item [3.] Poisson and Gamma regressions: The covariate vector $X$ is Gaussian or sub-Gaussian distributed, and the response $Y$ has a finite fourth moment.
        \end{enumerate}
        
\end{proposition}
\noindent \textit{Remark}:
\begin{itemize}
    \item [i)] We note that a random vector $X$ is said to have a finite $p$-th moment if $\mathbb{E}(\lVert X\rVert^p) < \infty$, where $\lVert \cdot\rVert^p$ is the $L^p$ norm.
    \item [ii)] Based on our derivations, we further observe that consistency is ensured if both the covariate vector and the response are supported on a compact set. This assumption is reasonable for many real-world problems, where the data typically falls within
a reasonable range.
\end{itemize}

Furthermore, we have the following result regarding a BIC-like criterion for regularized linear regression. Recall that the usual BIC criterion for linear regression is defined as
\begin{equation}\label{Eqs: BIC}
    \hat{K}_n = \min \argmin_{k \in \left[m\right]} \left\{R_{n}(\hat{\beta}_k) + \frac{\log(n)}{n} \hat{\text{df}}(\hat{\beta}_{k, n})\right\},
\end{equation}
where
\[\hat{\beta}_{k, n} = \argmin_{\beta \in \mathbb{T}_k} \,R_{n}(\beta)\]
and $\hat{\text{df}}(\hat{\beta}_{k, n})$ is the number of non-zero coefficients in $\hat{\beta}_{k, n}$, the unbiased estimator for the degrees of freedom established by \cite{zou_degrees_2007}. Details concerning this estimator can be found in Chapter 2.11 of \cite{ buhlmann_statistics_2011}. To the best of our knowledge, the consistency of the BIC has not been justified in the setting of regularized linear regression problems. Using the results from \cite{nguyen_panic:_2023}, we can establish the consistency of the following modifications of the BIC in (\ref{Eqs: BIC}):
\begin{equation}\label{Eqs: Modified BIC}
    \hat{K}_n = \min \argmin_{k \in \left[m\right]} \left\{R_{n}(\hat{\beta}_{k, n}) + \frac{\log(n)}{n} \left(\kappa\Tilde{\text{df}}(\hat{\beta}_{k, n}) + \varepsilon C_k\right)\right\}.
\end{equation}
Here, $\kappa$ is a non-negative constant and $\varepsilon$ is a small positive number. The term $\tilde{\text{df}}(\hat{\beta}_{k, n})$ is defined as follows: $\tilde{\text{df}}(\hat{\beta}_1) = \hat{\text{df}}(\hat{\beta}_1)$, and for $1 < k \in \left[m\right]$,
\[\Tilde{\text{df}}(\hat{\beta}_{k, n}) = \begin{cases}
    \hat{\text{df}}(\hat{\beta}_{k, n}), & \text{if } \hat{\text{df}}(\hat{\beta}_{k, n})\geq \hat{\text{df}}(\hat{\beta}_{j, n}),\, \forall j < k,\\
    \hat{\text{df}}(\hat{\beta}_{k-1, n}), & \text{otherwise}.
\end{cases}\]
To motivate these modifications, we note that the penalty shape of the BIC, denoted by $P^{\text{BIC}}_{k,n} = \hat{\text{df}}(\hat{\beta}_{k, n})$, is not strictly monotone and does not satisfy condition $\text{B2}^*$. In contrast, the modified penalty shape presented in (\ref{Eqs: Modified BIC}), given by $P_{k,n} = \kappa\Tilde{\text{df}}(\hat{\beta}_{k, n})+ \varepsilon C_k$, ensures strict monotonicity in accordance with $\text{B2}^*$, for any $\kappa \geq 0$ and $\varepsilon > 0$.

\begin{proposition}
    \label{Prop: BIC-like consistency}
    In the context of linear regression, let the covariate vector and response $(X, Y)$ be distributed on the measurable space $(\mathbb{R}^d \times \mathbb{R}, \mathcal{B}(\mathbb{R}^d)\otimes \mathcal{B}(\mathbb{R}))$ with probability measure $\Pi$. Assume both $X$ and $Y$ have a finite fourth moment, and the covariance matrix $\Sigma$ of $X$ is positive definite. That is, 
    \[\Sigma = \mathbb{E}\{(X - \mathbb{E}X)(X - \mathbb{E}X)^\top\}\]
    satisfies 
    \[x^\top \Sigma x > 0\]
    for any $x \in \mathbb{R}^d \backslash \{0\}$. Then, the criterion specified by (\ref{Eqs: Modified BIC}) is consistent. 
\end{proposition}

So far, we have formulated regression problems as model selection problems involving sets of finitely many models. A more natural formulation arises when we allow for an infinite number of candidate models, which occurs when the hypotheses are indexed by the entire positive real line or some compact interval within it, instead of an increasing sequence of numbers. This motivates the following result, which is given for certain sets of continuous and compact correspondences. Recall that a correspondence is a set-valued function. Specifically, a correspondence $\varphi$ from a set $\mathbb{X}$ to a set $\mathbb{Y}$ assigns to each $x\in \mathbb{X}$ a subset $\varphi(x)$ of $\mathbb{Y}$ \citep[Definition 17.1]{aliprantis_infinite_2006}.

\begin{theorem}\label{Thm: Infinite selection}
    Let $(X_i)_{i\in\left[n\right]}$ be an i.i.d sequence taking values in a set $\mathbb{X} \subset \mathbb{R}^d$ and $(\mathcal{H}_k)_{k\in\left[a, b\right]}$ be a set of hypothesis spaces of the form 
    \begin{align*}
        \mathcal{H}_k &= \{h(\cdot;\beta): \mathbb{X} \to \mathbb{R}: \beta \in \mathbb{T}_k\subset \mathbb{R}^d\} \text{,}
    \end{align*}
    where $a < b$. Let 
    $\mathcal{L}: k\mapsto \mathbb{T}_k$ be a continuous and compact-valued correspondence, and assume that the parameter spaces $(\mathbb{T}_k)_{k\in\left[a, b\right]}$ are nested. Assume that A1--A3, B1 and B2 hold for each $k$. Additionally, assume $P_{k,n}$ is a continuous and strictly increasing function in $k$ for all $n \in\mathbb{N}$. Define 
    \begin{align*}
        \mathcal{K} &= \argmin_{k \in \left[a, b\right]} \left\{\min_{\beta \in \mathbb{T}_k}r(\beta)\right\}, \quad k^* = \min_{k\in\mathcal{K}} k \text{.}
    \end{align*}
    Then, the PanIC estimator satisfies,
    \begin{equation}\label{Eqs: convergence in probability}
       \plim_{n\to\infty} \hat{K}_n = k^* \text{.}
    \end{equation}
\end{theorem}

In the context of regression problems, we shall define the parameter space $\mathbb{T}_k$ of the correspondence $\mathcal{L}: k \mapsto \mathbb{T}_k$ as 
\begin{equation}\label{Eqs: Parameter Space}
    \mathbb{T}_k = \begin{cases}
        \{(\beta_0, \beta) \, |\, \beta_0 \in \mathbb{U}, \beta\in \mathbb{R}^d: \lVert\beta\rVert_1 \leq k\}, & \text{for Lasso regression}\\
        \{(\beta_0, \beta) \, |\, \beta_0 \in \mathbb{U}, \beta\in \mathbb{R}^d: \lVert\beta\rVert_2 \leq k\}, & \text{for Ridge regression}\\
        \{(\beta_0, \beta) \, |\, \beta_0 \in \mathbb{U}, \beta\in \mathbb{R}^d: \alpha\lVert\beta\rVert_1 + (1-\alpha)\lVert\beta\rVert_2^2 \leq k\}, & \text{for Elastic-net regression}.
    \end{cases}
\end{equation}
When thusly defined, $\mathcal{L}: k \mapsto \mathbb{T}_k$ is compact-valued and continuous. Therefore, we have the following result.
\begin{corollary}
    Under the Assumptions of Theorem \ref{Thm: Infinite selection}, let $(\mathcal{H}_k)_{k\in\left[a, b\right]}$ be the set of hypothesis spaces for a regression problem. Specifically, the function $h$ is the mean function in linear, logistic, Poisson, or Gamma regression, as outlined in Table \ref{Table: Regression}, with the parameter spaces defined in  (\ref{Eqs: Parameter Space}). Suppose the corresponding conditions in Proposition \ref{Prop: Sufficient conditions} hold and $P_{k,n}$ is appropriately defined as in Theorem \ref{Thm: Infinite selection}. Then, the PanIC estimator is consistent in the sense of  (\ref{Eqs: convergence in probability}).
\end{corollary}

\subsection{Computation} \label{Sect: Computations}
In this section, we propose a computational method designed for implementing PanIC in regression problems within a finite sample setting. This is achieved via an application of duality in optimization theory, as covered, for example, in \cite{boyd_convex_2004}, \cite{kloft_lp-norm_2011}, and \cite{oneto_tikhonov_2016}. First, we review a useful result on duality. Suppose $f: \mathbb{R}^d \to \mathbb{R}$ is a convex function and $g: \mathbb{R}^d \to \mathbb{R}_{\geq 0}$ is a non-negative convex function. Consider the following constrained optimization problem:
\begin{align}\label{Eqs: General constrained problem}
\begin{split}
    \min \quad & \, f(\beta)\\
    \text{s.t.}\quad & \beta \in \mathbb{R}^d: g(\beta) \leq C \text{,}
\end{split}
\end{align}
for some value $C \in \mathbb{R}_+$. Assume that a solution exists and a constraint qualification holds, the solutions of (\ref{Eqs: General constrained problem}) are equivalent to that of its Lagrange dual problem \citep{kloft_lp-norm_2011}:
\begin{align}\label{Eqs: General unconstrained problem}
\begin{split}
    \min \quad & \, f(\beta) + \lambda g(\beta)\\
    \text{s.t.}\quad & \beta \in \mathbb{R}^d,
\end{split}
\end{align}
for some $\lambda \in \left[0, \infty\right)$, which is known as a regularization constant. Hence, for each $C$, there exists a $\lambda$ such that the optimal solutions of both problems coincide. The converse statement also holds, and if $f$ is strictly convex, the corresponding constrained problem is identified by the constraint $g(\beta)\leq g(\beta_\lambda)$, where $\beta_\lambda$ is the unique solution of (\ref{Eqs: General unconstrained problem}) \citep{kloft_lp-norm_2011}. This result allows us to define the function $\mathcal{L}:\lambda \mapsto g(\beta_\lambda)$, for all $\lambda \in \left[0, \infty\right)$. Under this setting, we can further establish the monotonicity and the continuity of the function $\mathcal{L}$. 
\begin{lemma}\label{Lemma: continuity}
    Suppose $f: \mathbb{R}^d \to \mathbb{R}_{\geq 0}$ is a strictly convex function and $g: \mathbb{R}^d \to \mathbb{R}_{\geq 0}$ is a non-negative convex function with a non-empty and bounded sub-level set $L_c = \{\beta\in \mathbb{R}^d\, |\, g(\beta) \leq c\}$, for some $c\in\mathbb{R}_{\geq 0}$. Then, the mapping $\mathcal{L}: \mathbb{R}_+ \to \mathbb{R}$, given by 
    \[\mathcal{L}(\lambda) = g(\beta_\lambda),\]
    is a continuous function, where
    \[ \beta_{\lambda}=\underset{\beta\in\mathbb{R}^{d}}{\arg\min}\ f\left(\beta\right)+\lambda g\left(\beta\right).
    \]
\end{lemma}

\begin{lemma}\label{Lemma: monotonicity} Let $\lambda_1$ and $\lambda_2$ be non-negative real numbers, with $\lambda_2 > \lambda_1$. Under the same assumptions as those of Lemma \ref{Lemma: continuity}, it holds that $g(\beta_{\lambda_2}) \leq g(\beta_{\lambda_1})$. Further, if $g(\beta_{\lambda_2}) < g(\beta_{\lambda_1})$, then $f(\beta_{\lambda_2}) > f( \beta_{\lambda_1})$. Else, if $g(\beta_{\lambda_2}) = g(\beta_{\lambda_1})$, then $\beta_{\lambda_2} = \beta_{\lambda_1}$. 
\end{lemma}

Our formulation of regression problems in the PanIC framework is associated with a sequence of constrained optimization problems, which can be expressed in the general form 
\begin{align}\label{Eqs: Constrained problem}
\begin{split}
    \min \quad & \, R_n(\beta_0, \beta) = \frac{1}{n} \sum^n_{i=1}\ell(x_i, y_i; \beta_0, \beta)\\
    \text{s.t.}\quad & \beta_0\in \mathbb{U},\, \beta\in \mathbb{R}^d: g(\beta) \leq C, \, C \in \mathbb{R}_+,
\end{split}
\end{align}
where $g$ is defined as the $l_1$ norm, the $l_2$ norm, or a combination of the two. Assuming the empirical risk function $R_n(\beta_0, \beta)$ is strictly convex, Lemma \ref{Lemma: continuity} and \ref{Lemma: monotonicity} allow us to use any root-finding algorithm, such as the bisection method, to find an (approximate) Lagrange dual problem of (\ref{Eqs: Constrained problem}). Compared to directly solving the constrained optimization problem, this approach greatly simplifies the computations. Algorithm \ref{Algo: PanIC Lasso} presents the computational steps for solving the Lasso problem. Generalizing this algorithm to both the Ridge regression and the Elastic-net regression is straightforward. 
\begin{algorithm}[h]
\caption{Computing the Lasso solution using PanIC}\label{Algo: PanIC Lasso}
\begin{algorithmic}[1]
\Require a sample of data $(x_i, y_i)_{i\in\left[n\right]}$, an increasing sequence $(C_k)_{k\in\left[m\right]}$, a function $h(\cdot; \beta_0, \beta): \mathcal{X}\to \mathcal{Y}$, a loss function $\ell: \mathcal{X}\times \mathcal{Y} \to \mathbb{R}$, a valid PanIC penalty function $P_{k,n}$.
\State Formulate a sequence of hypotheses $(\mathcal{H}_k)_{k \in \left[m\right]}$ with
\[\mathbb{T}_k \gets \{(\beta_0, \beta) \, |\beta_0\in \mathbb{R},\, \, \beta\in \mathbb{R}^d: \lVert\beta\rVert_1 \leq C_k\}\] 
and 
\[\mathcal{H}_k \gets \{h(\cdot; \beta_0, \beta)\,| (\beta_0, \beta) \in \mathbb{T}_k\}.\]
Associated with each hypothesis, define the optimization problem 
\begin{align}\label{Eqs: Constrained problem}
\begin{split}
    \argmin \quad & \, R_n(\beta_0, \beta) \\
    \text{s.t.}\quad & (\beta_0,\beta)\in \mathbb{T}_k.
\end{split}
\end{align}
\For{$k \in \left[m\right]$}
\State Determine the Lagrange dual problem of (\ref{Eqs: Constrained problem}) using a root-finding algorithm.
\State Denote $\hat{\vartheta}_k$ and $\hat{\beta}_k$ as the optimal value and the optimal solution of the Lagrange dual problem.
\EndFor
\State Compute the set of optimal hypotheses $\hat{\mathcal{K}}$, with
\[\hat{\mathcal{K}} \gets \argmin_{k \in \left[m\right]} \left\{\hat{\vartheta}_k + P_{k,n} \right\}.\]
\State Compute the PanIC estimate
\[\hat{K}_n \gets \min_{k \in \hat{\mathcal{K}}} \, k. \]
\Return $\hat{K}_n$, $\hat{\beta}_{\hat{K}_n}$
\end{algorithmic}
\end{algorithm}
% \footnotetext{$P_{k,n}$ must satisfy conditions B1 and B2}

\section{Simulation Study} \label{Sect: Simulation}

In this section, we present a set of simulated regression problems to assess the numerical performance of PanIC. We particularly focus on linear and logistic regression. 

\subsection{Simulation Setup}\label{Sect: Linear regression}
For each type of regression problem, we investigate three sample sizes: $n \in \{500, 1000, 2000\}$. The performance of PanIC on each sample size is simulated $N = 500$ times to obtain summary statistics. For each simulated regression problem, we employ Algorithm \ref{Algo: PanIC Lasso} to compute the PanIC, and then compare its performance to that of the 5-fold Cross-Validation (CV) and the modified BIC scheme, as described in \cite{stone_cross-validatory_1974} and (\ref{Eqs: Modified BIC}), respectively.

For the $j$-th regression problem, where $j \in \left[N\right]$, we simulate $n$ i.i.d random covariate vectors $(X_i^{(j)})_{i\in\left[n\right]}$ from the $20$-dimensional Gaussian distribution with zero mean and the identity covariance matrix, \textit{i.e.} $N(\boldsymbol{0}, \boldsymbol{I}_{20})$. Additionally, within the sparse learning framework, we generate a sparse true model vector $\beta^{*(j)}$ from $N(\boldsymbol{0}, \boldsymbol{I}_{20}^*)$, where 
\begin{align*}
    \boldsymbol{I}_{20}^* &= \begin{pmatrix}
    \boldsymbol{I}_{10} & \vline & \boldsymbol{0} \\
    \hline
    \boldsymbol{0} & \vline & \boldsymbol{0}\\
    \end{pmatrix}. 
\end{align*}
This distribution ensures $\beta^{*(j)}$ is a $20$-dimensional vector with $10$ active variables. For each observation $i$, where $i \in \left[ n\right]$, the response $Y_i^{(j)}$ is generated as a random variable following a normal distribution $N({\beta^{*(j)}}^\top X_i^{(j)}, \sigma^2)$, in the context of linear regressions, with $\sigma^2 > 0$. We further investigate three noise levels, given by $\sigma \in \{1, 2, 5\}$, to assess the performance of PanIC across different noise settings. For logistic regressions, the response is generated as a random variable following a Bernoulli distribution $\text{Bernoulli}(p^{(j)}_i)$, where 
\[p^{(j)}_i = f({\beta^{*(j)}}^\top X_i^{(j)})\]
and 
\[f(x) = \frac{1}{1 + \exp(-x)}.\]
To explore a range of model complexities, we construct an evenly spaced sequence $(C_k^{(j)})_{k\in \left[m\right]} \subset \mathbb{R}_+$, where $C_1^{(j)} = 0$, $C_m^{(j)} = \lVert \hat{\beta}^{(j)}\rVert_1$, and $\hat{\beta}^{(j)}$ is the solution obtained by minimizing the empirical risk function. 

As the final input for Algorithm \ref{Algo: PanIC Lasso}, we seek a valid penalty shape and a value of the hyperparameter $\kappa$, as identified in the form of (\ref{Eqs: Penalty Shape}). In our context, a natural choice for the penalty shape is 
\begin{equation}\label{Eqs: Norm Penalty Shape}
    \text{pen}_{\text{shape}}(k, n) =  C_k \sqrt{\frac{\log n}{n}}.
\end{equation}
Verifying that (\ref{Eqs: Norm Penalty Shape}) qualifies as a valid PanIC penalty is straightforward, as it satisfies conditions B1 and B2. 

\subsection{Hyperparameter Calibration for PanIC}\label{Sect: calibration}
The optimal calibration of $\kappa$ remains an open-ended task. To this end, we introduce several statistics, which help us evaluate a learning method's performance in finite sample settings. For each of the $N = 500$ simulation problems, we first employ an error function defined by
\[\text{error}(\hat{\beta}^{(j)}_{\hat{k}}) = \lVert\hat{\beta}^{(j)}_{\hat{k}}\rVert_1 - \lVert\beta^{*(j)}\rVert_1,\]
where $\hat{\beta}^{(j)}_{\hat{k}}$ denotes the chosen model of the $j$-th simulated problem, for $j\in\left[N\right]$. We use the signed value of this function to indicate the position of the selected model relative to the truth, and its absolute value to indicate the model's proximity to that truth. We note that this error function is of primary interest. We have established its asymptotic properties via the PanIC theory; however, its behavior in finite samples remains unknown. As complementary statistics, we report the number of active variables and the number of wrongly selected variables\footnote{A variable is considered wrongly selected in two situations: (i) its coefficient in $\beta^{*(j)}$ is non-zero, but its coefficient in $\hat{\beta}^{(j)}_{\hat{k}}$ is zero; (ii) its coefficient in $\hat{\beta}^{(j)}_{\hat{k}}$ is non-zero, but its coefficient in $\beta^{*(j)}$ is zero.} of the chosen model $\hat{\beta}^{(j)}_{\hat{k}}$, denoted as \#var and as \#w.var, respectively. Including these metrics enables a further understanding of PanIC's finite sample behavior, especially regarding its model selection capacity. However, we note that the theory for PanIC does not provide insights into the behavior of these metrics, whether in finite samples or asymptotically.    

\begin{figure}[H]
    \centering
    \includegraphics[scale=0.25]{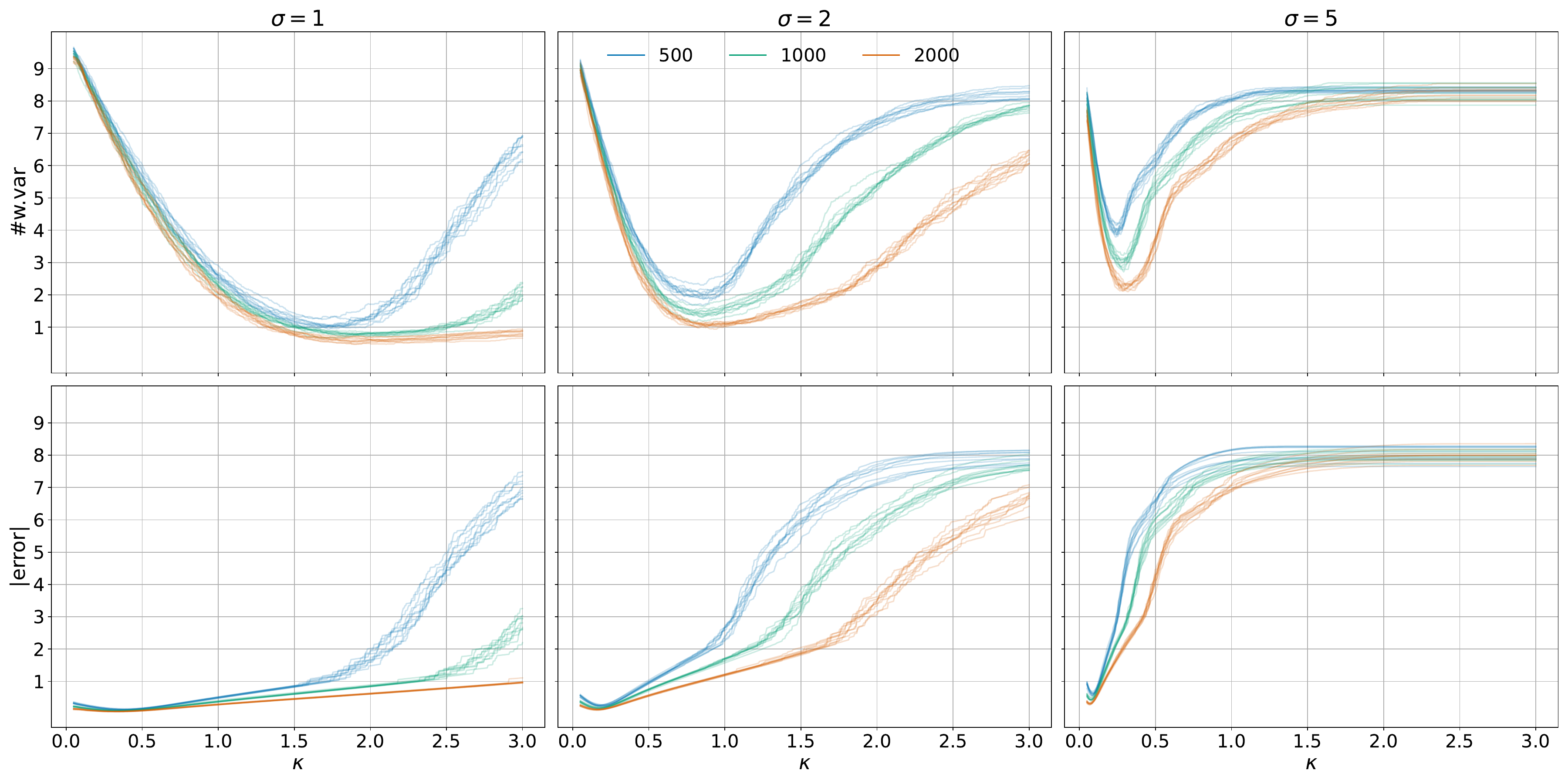}
    \caption{Performance of different $\kappa$ values in ten randomly selected linear regression problems of the $N = 500$ simulations.}
    \label{fig: w.var}
\end{figure}

We experiment with different values of $\kappa$ and record the preceding statistics of their learned model. The performance of each $\kappa$ in ten randomly selected regression problems is visualized in Figure \ref{fig: w.var}.

\subsection{Results and Discussion} \label{Sect: Discussion}
Concerning the calibration of
$\kappa$, we make the following observations. First, for our primary objective of estimating the norm of the optimal model, small values of $\kappa$ appear to be effective, as indicated by our simulation results. This is clear from the plots in the second row of Figure \ref{fig: w.var}, where it can be seen that small $\kappa$ values lead to a near-optimal value of the absolute error function. However, for model selection, this range of $\kappa$ values is inadequate. The plots in the first row of Figure \ref{fig: w.var} illustrate that models learned with small $\kappa$ values are dense, which is in direct contrast to the sparsity of the true model. 

In general, calibrating $\kappa$ for reasonable performance on \#w.var remains a challenging task. As shown in Figure \ref{fig: w.var}, the \#w.var statistic's sensitivity to $\kappa$ values varies significantly. For problems characterized by low noise, or a high signal-to-noise ratio, the learned model performs well across a broad range of $\kappa$ values. However, this robustness diminishes in high noise problems. For $\sigma = 5$, the range of $\kappa$ values that yield reasonable performance is significantly narrowed. Moreover, it appears impossible to identify a universal range of $\kappa$ values that performs well across different problem settings.

Tables \ref{Table: linear simulation error}--\ref{Table: logistic simulation} report summary statistics for models learned under PanIC, the modified BIC, and the CV schemes, in both linear and logistic regression scenarios. For PanIC, $\kappa$ is set to the value that yields optimal performance in \#w.var for each respective simulation setting, as determined by our calibration results. In the case of the modified BIC, $\kappa$ is fixed at $1$, on account of the similarity to the BIC (Eq. \ref{Eqs: BIC}) as shown in (\ref{Eqs: Modified BIC}). According to the theory for PanIC, both the PanIC and the modified BIC are asymptotically consistent in the linear setting, with PanIC also known to be consistent in the logistic setting. This aligns with the results reported in Tables \ref{Table: linear simulation error} and \ref{Table: logistic simulation error}. More interesting are the results in Tables \ref{Table: linear simulation} and \ref{Table: logistic simulation}, where we observe notable differences between the two schemes in their model selection capacity. It is evident that the modified BIC, denoted as $\text{BIC}_{\text{m}}$, is effective in selecting the correct model across all simulation settings. In comparison, while PanIC has the potential to be effective in model selection, this efficacy is conditional on the understanding of the problem, which enables prior calibration of $\kappa$.

\begin{figure}[H]
    \centering
    \includegraphics[scale=0.35]{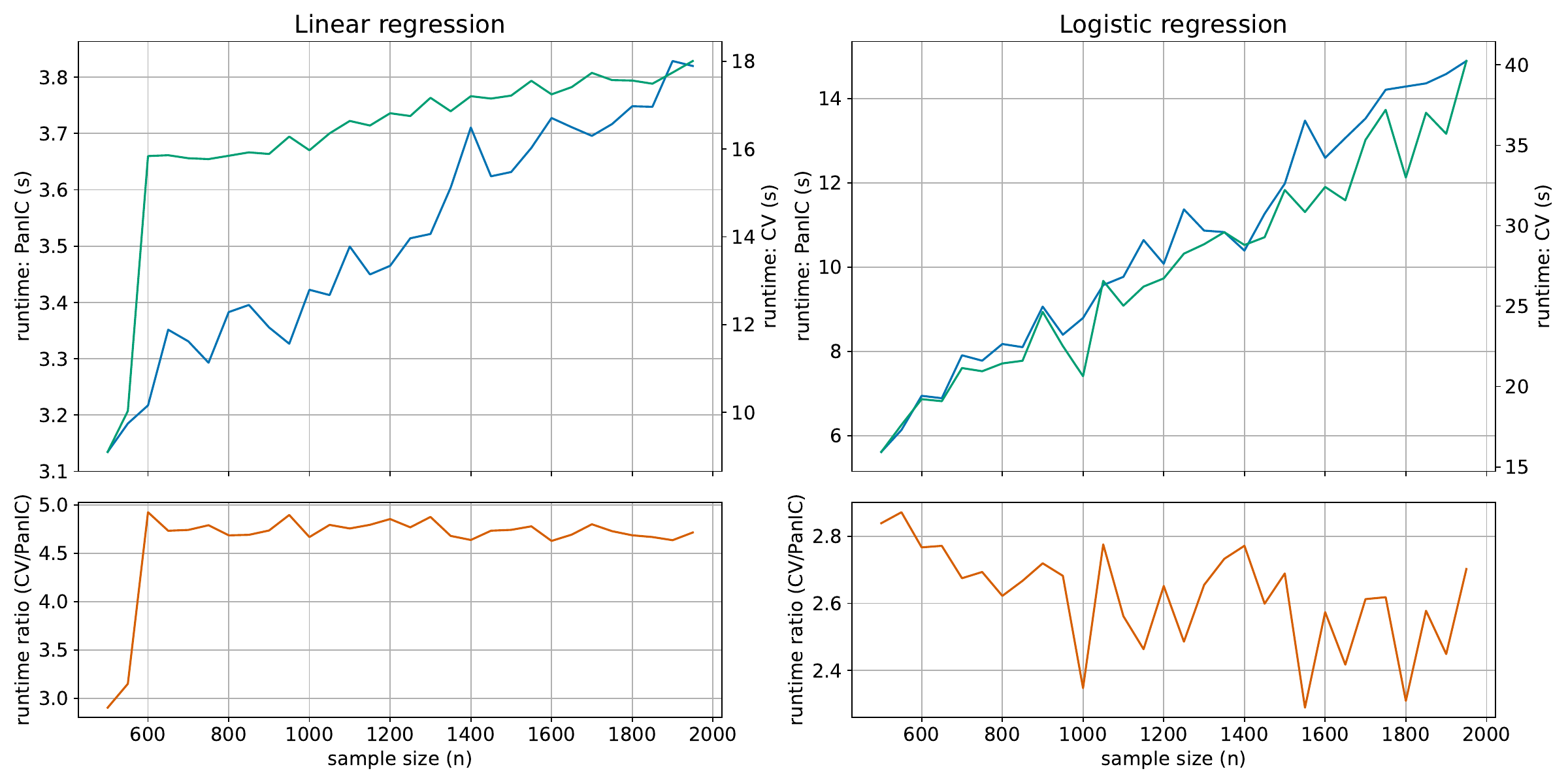}
    \caption{Total runtime of PanIC (in blue) and CV (in green) for linear and logistic regressions}
    \label{fig: profile}
\end{figure}

We note that both $\text{BIC}_{\text{m}}$ and PanIC are better suited for learning under sparsity than 5-fold CV. Tables \ref{Table: linear simulation} and \ref{Table: logistic simulation} indicate that specific $\kappa$ values for the modified BIC and PanIC lead to sparse solutions that closely approximate the true model. In contrast, CV consistently selects denser models. CV's ineffectiveness for learning under sparsity is briefly discussed by \cite{buhlmann_statistics_2011}, where it is noted that the CV scheme operates on optimizing prediction accuracy, which is often in conflict with variable selection. The goal of the latter is to recover the set of active variables, which often requires a sufficiently large regularization constant $\lambda$ that is not optimal for prediction. As a final remark on comparing PanIC with CV, our profiling demonstrates that PanIC is computationally more efficient than CV for sample sizes ranging from $500$ to $2000$, as visualized in Figure \ref{fig: profile}.

\begin{table}[H]
\caption{error and $\lvert\text{error}\rvert$ statistics of simulated linear regression problems}
\label{Table: linear simulation error}
\centering
\begin{adjustbox}{width=1\textwidth}
\small
\begin{tabular}{|c |c c |c c | c c |c c | c c |c c|} 
 \cline{2-13}
 \multicolumn{1}{c}{} &
  \multicolumn{4}{|c|}{$n = 500$} & \multicolumn{4}{c|}{$n = 1,000$} & \multicolumn{4}{c|}{$n = 2,000$}\\
 \multicolumn{1}{c}{} & \multicolumn{2}{|c}{error} & \multicolumn{2}{c|}{$\lvert\text{error}\rvert$} & \multicolumn{2}{c}{error} & \multicolumn{2}{c|}{$\lvert\text{error}\rvert$} & \multicolumn{2}{c}{error} & \multicolumn{2}{c|}{$\lvert\text{error}\rvert$} \rule{0pt}{0ex}\\
  \multicolumn{1}{c}{} & \multicolumn{1}{|c}{mean} & \multicolumn{1}{c}{se} & \multicolumn{1}{c}{mean} & \multicolumn{1}{c|}{se} &
  \multicolumn{1}{c}{mean} & \multicolumn{1}{c}{se} & \multicolumn{1}{c}{mean} & \multicolumn{1}{c|}{se} &
  \multicolumn{1}{c}{mean} & \multicolumn{1}{c}{se} & \multicolumn{1}{c}{mean} & \multicolumn{1}{c|}{se} \\
  [0.5ex] 
  \cline{2-13}
  \multicolumn{13}{c}{}
  \\[-0.9em]
  \hline
  % \hline
  % \hline
  \multicolumn{1}{|c|}{} &
  \multicolumn{12}{c|}{$\sigma = 1$} \\
  \hline
\text{PanIC} & -1.1907 & 0.0463 & 1.1907 & 0.0463 & -0.7581 & 0.0050 & 0.7581 & 0.0050 & -0.5598 & 0.0035 & 0.5598 & 0.0035 \\
$\text{BIC}_{\text{m}}$ & -0.4117 & 0.0104 & 0.4218 & -0.2884 & -0.2884 & 0.0079 & 0.3014 & 0.0068 & -0.2233 & 0.0053 & 0.2259 & 0.0051 \\
\text{CV} & 0.3012 & 0.0081 & 0.3061 & 0.0077 & 0.2308 & 0.0057 & 0.2332 & 0.0055 & 0.1647 & 0.0039 & 0.1674 & 0.0036 \\
 [0.5ex] 
  \hline\hline
\multicolumn{1}{|c|}{} &
  \multicolumn{12}{c|}{$\sigma = 2$} \\
  \hline
\text{PanIC} & -1.8612 & 0.0211 & 1.8612 & 0.0211 & -1.3461 & 0.0149 & 1.3461 & 0.0149 & -0.9931 & 0.0072 & 0.9931 & 0.0072 \\
$\text{BIC}_{\text{m}}$ & -0.8315 & 0.0224 & 0.8578 & 0.0203 & -0.5906 & 0.0159 & 0.6095 & 0.0144 & -0.4431 & 0.0103 & 0.4533 & 0.0094 \\
\text{CV} & 0.6648 & 0.0158 & 0.6707 & 0.0153 & 0.4607 & 0.0110 & 0.4673 & 0.0104 & 0.3221 & 0.0081 & 0.3283 & 0.0076 \\
 [0.5ex] 
  \hline\hline
\multicolumn{1}{|c|}{} &
  \multicolumn{12}{c|}{$\sigma = 5$} \\
  \hline
\text{PanIC} & -3.3285 & 0.0731 & 3.3285 & 0.0731 & -2.3476 & 0.0285 & 2.3476 & 0.0285 & -1.7429 & 0.0181 & 1.7429 & 0.0181 \\
$\text{BIC}_{\text{m}}$ & -2.3934 & 0.0544 & 2.4285 & 0.0512 & -1.6987 & 0.0391 & 1.7285 & 0.0364 & -1.1329 & 0.0268 & 1.1547 & 0.0248 \\
\text{CV} & 1.7180 & 0.0372 & 1.7246 & 0.0366 & 1.1857 & 0.0284 & 1.2049 & 0.0268 & 0.8681 & 0.0193 & 0.8743 & 0.0187 \\
 [0.5ex] 
  \hline
\end{tabular}
\end{adjustbox}
\end{table}

\begin{table}[H]
\caption{\#var and \#w.var statistics of simulated linear regression problems}
\label{Table: linear simulation}
\centering
\begin{adjustbox}{width=1\textwidth}
\small
\begin{tabular}{|c |c c |c c | c c |c c | c c |c c|} 
 \cline{2-13}
 \multicolumn{1}{c}{} &
  \multicolumn{4}{|c|}{$n = 500$} & \multicolumn{4}{c|}{$n = 1,000$} & \multicolumn{4}{c|}{$n = 2,000$}\\
 \multicolumn{1}{c}{} & \multicolumn{2}{|c}{\#var} & \multicolumn{2}{c|}{\#w.var} & \multicolumn{2}{c}{\#var} & \multicolumn{2}{c|}{\#w.var} & \multicolumn{2}{c}{\#var} & \multicolumn{2}{c|}{\#w.var} \rule{0pt}{0ex}\\
  \multicolumn{1}{c}{} & \multicolumn{1}{|c}{mean} & \multicolumn{1}{c}{se} & \multicolumn{1}{c}{mean} & \multicolumn{1}{c|}{se} &
  \multicolumn{1}{c}{mean} & \multicolumn{1}{c}{se} & \multicolumn{1}{c}{mean} & \multicolumn{1}{c|}{se} &
  \multicolumn{1}{c}{mean} & \multicolumn{1}{c}{se} & \multicolumn{1}{c}{mean} & \multicolumn{1}{c|}{se} \\
  [0.5ex] 
  \cline{2-13}
  \multicolumn{13}{c}{}
  \\[-0.9em]
  \hline
  % \hline
  % \hline
  \multicolumn{1}{|c|}{} &
  \multicolumn{12}{c|}{$\sigma = 1$} \\
  \hline
\text{PanIC} & 9.1580 & 0.0447 & 1.1340 & 0.0454 & 9.5140 & 0.0370 & 0.7900 & 0.0374 & 9.6380 & 0.0315 & 0.5500 & 0.0314 \\
$\text{BIC}_{\text{m}}$ & 11.3580 & 0.0832 & 2.4580 & 0.0665 & 11.3940 & 0.0799 & 2.1460 & 0.0676 & 11.3460 & 0.0690 & 1.8700 & 0.0631 \\
\text{CV} & 19.4640 & 0.0347 & 9.4960 & 0.0345 & 19.4820 & 0.0354 & 9.5260 & 0.0353 & 19.5620 & 0.0314 & 9.5900 & 0.0299 \\
 [0.5ex] 
  \hline\hline
\multicolumn{1}{|c|}{} &
  \multicolumn{12}{c|}{$\sigma = 2$} \\
  \hline
\text{PanIC} & 8.7480 & 0.0564 & 1.8720 & 0.0588 & 9.1280 & 0.0511 & 1.4800 & 0.0500 & 9.4840 & 0.0429 & 0.9560 & 0.0420 \\
$\text{BIC}_{\text{m}}$ & 10.6000 & 0.0938 & 2.7360 & 0.0707 & 10.8860 & 0.0877 & 2.3980 & 0.0672 & 11.0060 & 0.0725 & 1.9380 & 0.0631 \\
\text{CV} & 19.5020 & 0.0364 & 9.5820 & 0.0353 & 19.4880 & 0.0348 & 9.5360 & 0.0334 & 19.5020 & 0.0356 & 9.5460 & 0.0345 \\
 [0.5ex] 
  \hline\hline
\multicolumn{1}{|c|}{} &
  \multicolumn{12}{c|}{$\sigma = 5$} \\
  \hline
\text{PanIC} & 7.9860 & 0.0891 & 3.9460 & 0.0792 & 8.6940 & 0.0776 & 3.1620 & 0.0730 & 9.1360 & 0.0657 & 2.4080 & 0.0621 \\
$\text{BIC}_{\text{m}}$ & 7.8860 & 0.1183 & 4.0660 & 0.0790 & 8.7940 & 0.1138 & 3.4700 & 0.0732 & 9.6640 & 0.0941 & 2.7680 & 0.0632 \\
\text{CV} & 19.3720 & 0.0415 & 9.6000 & 0.0388 & 19.4540 & 0.0371 & 9.6260 & 0.0324 & 19.5260 & 0.0327 & 9.6300 & 0.0304 \\
 [0.5ex] 
  \hline
\end{tabular}
\end{adjustbox}
\end{table}

\begin{table}[H]
\caption{error and $\lvert\text{error}\rvert$ statistics of simulated logistic regression problems}
\label{Table: logistic simulation error}
\centering
\begin{adjustbox}{width=1\textwidth}
\small
\begin{tabular}{|c |c c |c c | c c |c c | c c |c c|} 
 \cline{2-13}
 \multicolumn{1}{c}{} &
  \multicolumn{4}{|c|}{$n = 500$} & \multicolumn{4}{c|}{$n = 1,000$} & \multicolumn{4}{c|}{$n = 2,000$}\\
 \multicolumn{1}{c}{} & \multicolumn{2}{|c}{error} & \multicolumn{2}{c|}{$\lvert\text{error}\rvert$} & \multicolumn{2}{c}{error} & \multicolumn{2}{c|}{$\lvert\text{error}\rvert$} & \multicolumn{2}{c}{error} & \multicolumn{2}{c|}{$\lvert\text{error}\rvert$} \rule{0pt}{0ex}\\
  \multicolumn{1}{c}{} & \multicolumn{1}{|c}{mean} & \multicolumn{1}{c}{se} & \multicolumn{1}{c}{mean} & \multicolumn{1}{c|}{se} &
  \multicolumn{1}{c}{mean} & \multicolumn{1}{c}{se} & \multicolumn{1}{c}{mean} & \multicolumn{1}{c|}{se} &
  \multicolumn{1}{c}{mean} & \multicolumn{1}{c}{se} & \multicolumn{1}{c}{mean} & \multicolumn{1}{c|}{se} \\
  [0.5ex] 
  \cline{2-13}
  \multicolumn{13}{c}{}
  \\[-0.9em]
  \hline
  % \hline
  % \hline
\text{PanIC} & -4.6068 & 0.0610 & 4.6068 & 0.0610 & -3.9045 & 0.0534 & 3.9045 & 0.0534 & -3.2755 & 0.0501 & 3.2755 & 0.0501 \\
$\text{BIC}_{\text{m}}$ & -3.5468 & 0.0560 & 3.5560 & 0.0548 & -2.5519 & 0.0436 & 2.5540 & 0.0433 & -1.9428 & 0.0341 & 1.9428 & 0.0341 \\
\text{CV} & 1.5344 & 0.0478 & 1.5677 & 0.0456 & 0.9107 & 0.0298 & 0.9514 & 0.0272 & 0.5996 & 0.0197 & 0.6288 & 0.0177 \\
 [0.5ex] 
  \hline
\end{tabular}
\end{adjustbox}
\end{table}

\begin{table}[H]
\caption{\#var and \#w.var statistics of simulated logistic regression problems}
\label{Table: logistic simulation}
\centering
\begin{adjustbox}{width=1\textwidth}
\small
\begin{tabular}{|c |c c |c c | c c |c c | c c |c c|} 
 \cline{2-13}
 \multicolumn{1}{c}{} &
  \multicolumn{4}{|c|}{$n = 500$} & \multicolumn{4}{c|}{$n = 1,000$} & \multicolumn{4}{c|}{$n = 2,000$}\\
 \multicolumn{1}{c}{} & \multicolumn{2}{|c}{\#var} & \multicolumn{2}{c|}{\#w.var} & \multicolumn{2}{c}{\#var} & \multicolumn{2}{c|}{\#w.var} & \multicolumn{2}{c}{\#var} & \multicolumn{2}{c|}{\#w.var} \rule{0pt}{0ex}\\
  \multicolumn{1}{c}{} & \multicolumn{1}{|c}{mean} & \multicolumn{1}{c}{se} & \multicolumn{1}{c}{mean} & \multicolumn{1}{c|}{se} &
  \multicolumn{1}{c}{mean} & \multicolumn{1}{c}{se} & \multicolumn{1}{c}{mean} & \multicolumn{1}{c|}{se} &
  \multicolumn{1}{c}{mean} & \multicolumn{1}{c}{se} & \multicolumn{1}{c}{mean} & \multicolumn{1}{c|}{se} \\
  [0.5ex] 
  \cline{2-13}
  \multicolumn{13}{c}{}
  \\[-0.9em]
  \hline
  % \hline
  % \hline
\text{PanIC} & 8.3020 & 0.0603 & 2.6260 & 0.0613 & 8.6620 & 0.0543 & 1.9420 & 0.0555 & 9.0440 & 0.0469 & 1.4080 & 0.0474 \\
$\text{BIC}_{\text{m}}$ & 8.6020 & 0.0985 & 2.9940 & 0.0661 & 9.6320 & 0.0852 & 2.3760 & 0.0602 & 10.2840 & 0.0793 & 2.0840 & 0.0602 \\
\text{CV} & 19.5940 & 0.0283 & 9.7060 & 0.0288 & 19.6300 & 0.0292 & 9.6980 & 0.0277 & 19.6640 & 0.0263 & 9.6920 & 0.0258 \\
 [0.5ex] 
  \hline
\end{tabular}
\end{adjustbox}
\end{table}

\section{Conclusion}\label{Sect: Conclusion}
In this work, we focus on consistently estimating regression problems using a class of information criteria known as PanIC for regression problems, including linear, logistic, Poisson, and Gamma regressions. In addition to theoretical discussions, we present a simulation study of PanIC in the finite sample setting. Specifically, the results of our simulations indicate that PanIC's performance is comparable to that of cross-validation and BIC. Extending from our consistency result on sets of finitely many models, we introduce a new result (Theorem \ref{Thm: Infinite selection}) to consistently estimate over certain sets of uncountably infinitely many models. The setting adopted in Theorem \ref{Thm: Infinite selection} is natural for regressions, but can be extended to a broader scope.

We observe two interesting extensions of this research topic. First, we are interested in establishing consistency when the models are identified by random parameter spaces. For example, in a regression setting it is often useful to formulate a sequence of model spaces $(\mathcal{H}_k)$ given by $\mathcal{H}_k = \{h(\cdot;\beta): \mathbb{X} \to \mathbb{R}: \beta \in \mathbb{T}_k\}$ and $\mathbb{T}_k = \{\beta\in \mathbb{R}^d: \lVert\beta\rVert_1 \leq C_k(\omega)\}$, where $C_k(\omega)$ is some random variable that depends on the data. This formulation corresponds to solving regression problems as a sequence of regularization problems, where the sequence of regularization constants $(\lambda_k)$ induces a data-dependent restriction on the parameter spaces $(\mathbb{T}_k)$. For general learning problems that involve regularization, consistency under this setting allows us to directly estimate these problems in their unconstrained form, which is often useful. Secondly, we are interested in exploring the finite sample property of a PanIC estimator in greater detail. Further investigation may focus on the calibration of the unknown multiplicative factor $\kappa$ of a PanIC penalty term. 

\section*{Acknowledgements}
The authors acknowledge funding from the Australian Research Council grant DP2301009.

\appendix

\section{Proofs}
\subsection{Proof of Theorem \ref{Thm: Infinite selection}}
\begin{proof}
    Since $\mathcal{L}$ is a compact-valued continuous correspondence, by the Berge Maximum Theorem \citep[Theorem 17.31]{aliprantis_infinite_2006}, the value function defined by $m(k) = \min_{\beta \in \mathbb{T}_k}r(\beta)$ is continuous. A continuous function attains its minimum on a compact set, so $\mathcal{K}$ is non-empty. As the expected risk function $r$ is continuous, the set $\mathcal{K}$ contains a minimum value and $k^*$ is well-defined. By a similar argument, for each $n \in\mathbb{N}$, $\hat{K}_n$ is well-defined. To show convergence in probability, we will show for all $\sigma > 0$, $\mathbb{P}\left(\left|\hat{K}_n - k^*\right| \geq \sigma\right) \to 0$ as $n\to\infty$. Fix $\sigma > 0$, consider the case $\hat{K}_n \leq k^* -\sigma$. If $\sigma > k^* - a$, the statement is trivial. So assume $\sigma \leq k^* - a$. Let $\varepsilon$ be
    \[\varepsilon = \frac{1}{2} \left\{\min_{\beta \in \mathbb{T}_{k^*-\sigma}} r(\beta) - \min_{\beta \in \mathbb{T}_{k^*}} r(\beta) \right\}.\]
    Conditions A1--A3 and the i.i.d assumption are sufficient for the uniform strong law of large numbers \citet[Thm. 9.60]{shapiro_lectures_2021}, which gives $R_{n}(\beta) \to r(\beta)$ uniformly on any compact set $\mathbb{T}_k$ with probability one as $n\to\infty$. Using this result, and assumption in B1, for every $\delta > 0$, there exists some $N_\delta\in\mathbb{N}$ such that for all $n> N_\delta$, the events
    \begin{align*}
        \left|\min_{\beta \in \mathbb{T}_{k^*-\sigma}} R_{n} (\beta) - \min_{\beta \in \mathbb{T}_{k^*-\sigma}} r(\beta)\right| &\leq \frac{\varepsilon}{3},\\
        \left|\min_{\beta \in \mathbb{T}_{k^*}} R_{ n} (\beta) - \min_{\beta \in \mathbb{T}_{k^*}} r(\beta)\right| &\leq \frac{\varepsilon}{3},
    \end{align*}
    and $P_{k^*, n} \leq \varepsilon/3$ occur with probability at least $1 - \delta$. These events imply the following
    \begin{align*}
        \min_{\beta \in \mathbb{T}_{k^*-\sigma}} R_{n}(\beta) &\geq \min_{\beta \in \mathbb{T}_{k^*-\sigma}} r(\beta) - \frac{\varepsilon}{3}\\
        &= \min_{\beta \in \mathbb{T}_{k^*}} r(\beta) + \frac{5\varepsilon}{3}\\
        &\geq \min_{\beta \in \mathbb{T}_{k^*}} R_{n}(\beta) + P_{k^*, n} + \varepsilon.
    \end{align*}
    So for all $k \in \left[a, k^* - \sigma\right]$, 
    \begin{align*}
        \min_{\beta \in \mathbb{T}_{k^*}} R_{n}(\beta) + P_{k^*, n} &< \min_{\beta \in \mathbb{T}_{k^*-\sigma}} R_{n}(\beta) + P_{a, n}\\
        &\leq \min_{\beta \in \mathbb{T}_k} R_{n}(\beta) + P_{a, n}\\
        &\leq \min_{\beta \in \mathbb{T}_k} R_{n}(\beta) + P_{k, n}
    \end{align*}
    occurs with probability at least $1 - \delta$. Take $n\to\infty$, we have $\mathbb{P}\left(\hat{K}_n \leq k^*- \sigma \right) \to 0$. Now consider the case $\hat{K}_n \geq k^* + \sigma$. If $\sigma > b - k^*$, the statement is trivial. So assume $\sigma \leq b - k^*$.
    Since $\mathbb{T}_b$ is compact, \cite[Thm. 5.7]{shapiro_lectures_2021} implies that 
    \[\sqrt{n}\left(\min_{\beta \in \mathbb{T}_b} R_{n}(\beta) - \min_{\beta \in \mathbb{T}_b} r(\beta) \right) = O_{\mathbb{P}}(1).\]
    Since 
    \[\min_{\beta \in \mathbb{T}_{k^*}} r(\beta) = \min_{\beta \in \mathbb{T}_{b}} r(\beta), \]
    we have 
    \[\sqrt{n}\left(\min_{\beta \in \mathbb{T}_{b}} R_{n}(\beta) - \min_{\beta \in \mathbb{T}_{k^*}} R_{n}(\beta) \right) = O_{\mathbb{P}}(1).\]
    Fix $\delta > 0$, there exists $M>0$ such that,
    \begin{align*}
        \sqrt{n}\left|\min_{\beta \in \mathbb{T}_{b}} R_{n}(\beta) - \min_{\beta\in\mathbb{T}_{k^*}} R_{n}(\beta)\right| &\leq M
    \end{align*}
    with probability at least $1 - \delta$ for large $n$. By B2, for each $\delta > 0$ and $M > 0$,
    \[\sqrt{n}\{P_{k^*+\sigma,n} - P_{k^*, n}\} > M\]
    with probability at least $1-\delta$ for large $n$. So with probability $1 - 2\delta$, we have
    \[\min_{\beta\in\mathbb{T}_{k^*}} R_{n}(\beta) - \min_{\beta\in\mathbb{T}_{b}} R_{n}(\beta)\leq \frac{M}{\sqrt{n}} < P_{k^*+\sigma,n} - P_{k^*,n}\]
    and thus for all $k\in \left[k^*+\sigma, b\right]$,
    \begin{align*}
        \min_{\beta\in\mathbb{T}_{k^*}} R_{n}(\beta) + P_{k^*, n} &< \min_{\beta\in\mathbb{T}_{b}} R_{n}(\beta) + P_{k^*+\sigma, n}\\
        &\leq  \min_{\beta\in\mathbb{T}_{k}} R_{n}(\beta) + P_{k^*+\sigma, n}\\
        &\leq  \min_{\beta\in\mathbb{T}_{k}} R_{n}(\beta) + P_{k, n}.
    \end{align*}
    So $\lim_{n\to\infty} \mathbb{P}(\hat{K}_n - k^* \geq \sigma) = 0$ for all $\sigma > 0$.
\end{proof}

\subsection{Proof of Proposition \ref{Prop: Sufficient conditions}}
For simplicity, we show the results under the setting of Lasso regression, with a simplified mean function $h(x;\beta) = \beta^\top x$ (without bias term). Upon examining the argument, it is clear that these results are also applicable to both Ridge and Elastic-net regressions, and with the complete mean function $h(x;\beta_0, \beta) = \beta_0 + \beta^\top x$. 
\subsubsection{Linear Regression}
\begin{proof}
    It suffices to verify the PanIC conditions, 
    \begin{enumerate}
    \item[A1]: $\ell_k(x, y; \cdot): \mathbb{T}_k \to \mathbb{R}$ is continuous for each $(x, y) \in \mathbb{X}$ as it is differentiable with respect to $\beta$. Fix $\beta \in \mathbb{T}_k$, $\ell_k(\cdot; \beta): \mathbb{X} \to \mathbb{R}$ is given by
    \begin{align*}
        \ell_k(x, y; \beta) &= (\Tilde{f}(x, y; \beta)  - \Tilde{g}(x, y; \beta))^2\\
        &= (f(y; \beta)  - g(x; \beta))^2,
    \end{align*}
    where $\Tilde{f}(x, y; \beta) = f(y; \beta) = y$ and $\Tilde{g}(x, y; \beta) = g(x; \beta) = \beta^\top x$. The functions $f$ and $g$ are measurable on their respective domain, so $\Tilde{f}$ and $\Tilde{g}$ are measurable on the product space $\mathbb{X}$. It follows that $\ell_k(\cdot; \beta)$ is measurable on $\mathbb{X}$. 
    \item[A2]: $\mathbb{T}_k = \{\beta\in \mathbb{R}^d: \lVert\beta\rVert_1 \leq C_k\}$ is compact. Let $\tau_k = 0 \in \mathbb{R}^d$, as $\lVert 0 \rVert_1 \leq C_k$ for all $k \in \left[m\right]$, $\tau_k \in \mathbb{T}_k$. This condition is satisfied if $Y$ has a finite fourth moment:
    \[\mathbb{E}\left(\ell_k(X, Y; \tau_k)^2\right) = \mathbb{E}(Y^4) < \infty.\]
    \item[A3]: Fix $(x, y)\in \mathbb{X}$, $k \in \left[m\right]$ and $\beta, \beta' \in \mathbb{T}_k$. By the multivariate Mean Value Theorem \citep{davidson_stochastic_2021},
    \[|\ell_k(x, y; \beta) - \ell_k(x, y; \beta')| \leq \sup_{\beta^*} \left\lVert \left. \frac{\partial \ell_k}{\partial \beta}\right|_{\beta = \beta^*} \right \rVert_2 \lVert \beta - \beta'\rVert_2, \]
    where $\beta^*$ is any point on the line segment joining $\beta$ and $\beta'$. We can write $\beta^* = \lambda\beta + (1 - \lambda)\beta'$, for $\lambda\in\left[0, 1\right]$. Note that,
    \begin{align*}
        \left\lVert \left. \frac{\partial \ell_k}{\partial \beta}\right|_{\beta = \beta} \right \rVert_2 &= \lVert-2(y - \beta^\top x)x\rVert_2\\
        &= 2 \left\lvert y -  \beta^\top x\right\rvert\lVert x\rVert_2.
    \end{align*}
    This gives us
    \begin{align*}
        \sup_{\beta^*} \left\lVert \left. \frac{\partial \ell_k}{\partial \beta}\right|_{\beta = \beta^*} \right \rVert_2 &= \sup_{\lambda \in \left[0,1\right]} 2 \left\lvert y - (\lambda \beta  + (1 - \lambda)\beta')^\top x \right\rvert\lVert x\rVert_2.
    \end{align*}
    Using triangle inequality, we achieve the bound
        \begin{align*}
        \sup_{\beta^*} \left\lVert \left. \frac{\partial \ell_k}{\partial \beta}\right|_{\beta = \beta^*} \right \rVert_2 &\leq \sup_{\lambda \in \left[0,1\right]} 2 \left(|y| + \lambda \left|\beta^\top x\right| + (1 - \lambda) \left|\beta'^\top x\right|\right)\lVert x\rVert_2\\
        &\leq \sup_{\lambda \in \left[0,1\right]} 2 \left(|y| + \lambda \left\lVert x\right\rVert_1 \lVert\beta\rVert_1 + (1 - \lambda) \left\lVert x\right\rVert_1 \lVert\beta'\rVert_1\right)\lVert x\rVert_2\\
        & \leq 2 \left(|y| + \left\lVert x\right\rVert_1 \lVert\beta\rVert_1 + \left\lVert x\right\rVert_1 \lVert\beta'\rVert_1\right)\lVert x\rVert_2\\
        & \leq 2 \left(|y| + \left\lVert x\right\rVert_1 \lVert\beta\rVert_1 + \left\lVert x\right\rVert_1 \lVert\beta'\rVert_1\right)\lVert x\rVert_1\\
        &\leq 2 \left(|y| + 2\left\lVert x\right\rVert_1 C_k\right)\lVert x\rVert_1.
    \end{align*}
    Let $\mathcal{G}_k(x, y) = 2 \left(|y| + 2\left\lVert x\right\rVert_1 C_k\right)\lVert x\rVert_1$, then the condition $\mathbb{E}(\mathcal{G}_k(X, Y)^2) < \infty$ is satisfied if both $X$ and $Y$ have a finite fourth moment.
\end{enumerate}
\end{proof}

\subsubsection{Logistic Regression}
\begin{proof} $ $
    \begin{enumerate}
    \item[A1]: $\ell_k(x, y; \cdot): \mathbb{T}_k \to \mathbb{R}$ is continuous for each $(x, y) \in \mathbb{X}$ as it is differentiable with respect to $\beta$. Fix $\beta \in \mathbb{T}_k$, $\ell_k(\cdot; \beta): \mathbb{X} \to \mathbb{R}$ is given by
    \begin{align*}
        \ell_k(x, y; \beta) &= \Tilde{f}(x, y; \beta)\Tilde{g}(x, y; \beta)  + \Tilde{h}(x, y; \beta)\\
        &= f(y; \beta)g(x; \beta)  + h(x; \beta), 
    \end{align*}
    where $\Tilde{f}(x, y; \beta) = f(y; \beta) = y$,  
    \begin{align*}
        \Tilde{g}(x, y; \beta) = g(x; \beta) &= -\log\left(\frac{1}{1 + e^{-\beta^\top x}}\right) + \log\left(1 - \frac{1}{1 + e^{-\beta^\top x}}\right),\\
        \Tilde{h}(x, y; \beta) = h(x; \beta) &= -\log\left(1 - \frac{1}{1 + e^{-\beta^\top x}}\right). 
    \end{align*}
    Both $g$ and $h$ are continuous functions, so they are measurable with respect to the Borel $\sigma$-algebra on $\mathbb{R}^d$. The function $f$ is measurable with respect to the discrete $\sigma$-algebra on $\{0, 1\}$. So  $\Tilde{f}$, $\Tilde{g}$ and $\Tilde{h}$ are measurable on the product space $\mathbb{X}$. It follows that $\ell_k(\cdot; \beta)$ is measurable on $\mathbb{X}$. 
    \item[A2]: $\mathbb{T}_k$ is compact. Let $\tau_k = 0 \in \mathbb{R}^d$, then $\tau_k \in \mathbb{T}_k$ for all $k \in \left[m\right]$. This condition is satisfied as
    \begin{align*}
        \mathbb{E}\left(\ell_k(X, Y; \tau_k)^2\right) &= \mathbb{E}\left(\left(-Y \log(1/2) - (1-Y)\log(1-1/2)\right)^2\right)\\
        &= \mathbb{E}\left(\left(-\log(1/2)\right)^2\right)\\
        &= (\log(1/2))^2 < \infty.
    \end{align*}
    \item[A3]: Fix $(x, y)\in \mathbb{X}$, $k \in \left[m\right]$ and $\beta, \beta'\in \mathbb{T}_k$. By the multivariate Mean Value Theorem, we have
    \[|\ell_k(x, y; \beta) - \ell_k(x, y; \beta')| \leq \sup_{\beta^*} \left\lVert \left. \frac{\partial \ell_k}{\partial \beta}\right|_{\beta = \beta^*} \right \rVert_2 \lVert \beta - \beta'\rVert_2, \]
    where $\beta^*$ is any point on the line segment joining $\beta$ and $\beta'$. We again write $\beta^* = \lambda\beta + (1 - \lambda)\beta'$, for $\lambda\in\left[0, 1\right]$. Note that,
    \begin{align*}
        \left.\frac{\partial \ell_k}{\partial \beta}\right|_{\beta = \beta} = -\left(\frac{e^{-\beta^\top x}}{1 + e^{-\beta^\top x}}y - \frac{e^{-\beta^\top x}(1-y)}{(1+e^{-\beta^\top x})^2(1-(1+e^{-\beta^\top x})^{-1})}\right)x.
    \end{align*}
    Then, we have
    \begin{align*}
        \left\lVert \left. \frac{\partial \ell_k}{\partial \beta}\right|_{\beta = \beta} \right \rVert_2 &= \left\lVert-\left(\frac{e^{-\beta^\top x}}{1 + e^{-\beta^\top x}}y - \frac{e^{-\beta^\top x}(1-y)}{(1+e^{-\beta^\top x})^2(1-(1+e^{-\beta^\top x})^{-1})}\right)x\right\rVert_2\\
        &= \left\lVert\frac{e^{-\beta^\top x}}{1 + e^{-\beta^\top x}}y - \frac{e^{-\beta^\top x}(1-y)}{(1+e^{-\beta^\top x})^2(1-(1+e^{-\beta^\top x})^{-1})}\right\rVert_2\left\lVert x\right\rVert_2\\
        &\leq \left(\left\lVert\frac{e^{-\beta^\top x}}{1 + e^{-\beta^\top x}}\right\rVert_2 + \left\lVert\frac{e^{-\beta^\top x}}{(1+e^{-\beta^\top x})^2 - (1+e^{-\beta^\top x})}\right\rVert_2\right)\left\lVert x\right\rVert_2\\
        &\leq 2\left\lVert x\right\rVert_2\\
        &\leq 2\left\lVert x\right\rVert_1.
    \end{align*}
    It follows that
    \begin{align*}
        \sup_{\beta^*} \left\lVert \left. \frac{\partial \ell_k}{\partial \beta}\right|_{\beta = \beta^*} \right \rVert_2 &\leq 2\left\lVert x\right\rVert_1.
    \end{align*}
    Let $\mathcal{G}_k(x, y) = 2\lVert x\rVert_1$, then the condition $\mathbb{E}(\mathcal{G}_k(X, Y)^2) < \infty$ is satisfied if $X$ has a finite second moment.
\end{enumerate}
\end{proof}

\subsubsection{Poisson Regression}
\begin{proof} $ $
    \begin{enumerate}
    \item[A1]: $\ell_k(x, y; \cdot): \mathbb{T}_k \to \mathbb{R}$ is continuous for each $(x, y) \in \mathbb{X}$ as it is differentiable with respect to $\beta$. Fix $\beta \in \mathbb{T}_k$, $\ell_k(\cdot; \beta): \mathbb{X} \to \mathbb{R}$ is given by
    \begin{align*}
        \ell_k(x, y; \beta) &= \Tilde{f}(x, y; \beta)\Tilde{g}(x, y; \beta)  + \Tilde{h}_1(x, y; \beta) + \Tilde{h}_2(x, y; \beta)\\
        &= f(y; \beta)g(x; \beta)  + h_1(x; \beta) + h_2(y; \beta),
    \end{align*}
    where $\Tilde{f}(x, y; \beta) = f(y; \beta) = y$, 
    \begin{align*}
        \Tilde{g}(x, y; \beta) = g(x; \beta) &= -\beta^\top x,\\
        \Tilde{h}_1(x, y; \beta) = h_1(x; \beta) &= \exp(\beta^\top x),\\
        \Tilde{h}_2(x, y; \beta) = h_2(y; \beta) &= \log(y!).
    \end{align*}
    Both $g$ and $h_1$ are continuous functions, so they are measurable with respect to the Borel $\sigma$-algebra on $\mathbb{R}^d$. Both $f$ and $h_2$ are measurable with respect to the discrete $\sigma$-algebra on $\mathbb{Z}_{\geq 0}$. So  $\Tilde{f}$, $\Tilde{g}$, $\Tilde{h}_1$ and $\Tilde{h}_2$ are measurable on the product space $\mathbb{X}$. It follows that $\ell_k(\cdot; \beta)$ is measurable on $\mathbb{X}$. 
    
    \item[A2]: $\mathbb{T}_k$ is compact. Let $\tau_k = 0 \in \mathbb{R}^d$, then $\tau_k \in \mathbb{T}_k$ for all $k \in \left[m\right]$. This condition is satisfied if $Y$ has a finite fourth moment 
    \begin{align*}
        \mathbb{E}\left(\ell_k(X, Y; \tau_k)^2\right) &= \mathbb{E}\left(\left(1+\log(Y!)\right)^2\right)\\
        &\leq \mathbb{E}\left(\left(1+ Y\log(Y)\right)^2\right)\\
        &\leq \mathbb{E}\left(\left(1+ Y^2\right)^2\right)\\
        &= \mathbb{E}\left(Y^4 + 2Y^2 + 1\right) < \infty.
    \end{align*}
    \item[A3]: Fix $(x, y)\in \mathbb{X}$, $k \in \left[m\right]$ and $\beta, \beta'\in \mathbb{T}_k$. By the multivariate Mean Value Theorem,
    \[|\ell_k(x, y; \beta) - \ell_k(x, y; \beta')| \leq \sup_{\beta^*} \left\lVert \left. \frac{\partial \ell_k}{\partial \beta}\right|_{\beta = \beta^*} \right \rVert_2 \lVert \beta - \beta'\rVert_2,\]
    where $\beta^*$ is any point on the line segment joining $\beta$ and $\beta'$. We write $\beta^* = \lambda\beta + (1 - \lambda)\beta'$, for $\lambda\in\left[0, 1\right]$. Note that,
    \begin{align*}
        \left.\frac{\partial \ell_k}{\partial \beta}\right|_{\beta=\beta} = \left(\exp( \beta^\top x) - y\right)x.
    \end{align*}
    Using triangle inequality,
    \begin{align*}
        \sup_{\beta^*} \left\lVert \left. \frac{\partial \ell_k}{\partial \beta}\right|_{\beta = \beta^*} \right \rVert_2 &= \sup_{\lambda\in\left[0, 1\right]}\left| \exp\left((\lambda\beta + (1 - \lambda)\beta')^\top x\right) - y\right| \lVert x\rVert_2\\
        &\leq \left| \exp\left(\lvert\beta^\top x\rvert + \lvert\beta'^\top x\rvert\right) - y\right| \lVert x\rVert_2\\
        &\leq \left( \exp\left(\lvert\beta^\top x\rvert + \lvert\beta'^\top x\rvert\right) + \left|y\right|\right) \lVert x\rVert_2\\
        &\leq \left( \exp(\lVert\beta\rVert_1\lVert x\rVert_1 + \lVert\beta'\rVert_1\lVert x\rVert_1) + \left|y\right|\right) \lVert x\rVert_2\\
        &\leq \left( \exp(2C_k\lVert x\rVert_1) + y\right) \lVert x\rVert_1.
    \end{align*}
    Let $\mathcal{G}_k(x, y) = \left( \exp(2C_k\lVert x\rVert_1) + y\right) \lVert x\rVert_1$, then the condition $\mathbb{E}(\mathcal{G}_k(X, Y)^2) < \infty$ is satisfied if 
    \begin{align}\label{Eqs: Poisson condition}
        \mathbb{E}\left(\mathcal{G}_k(X, Y)^2\right) &= \mathbb{E}\left(\left( \exp(2C_k\lVert X\rVert_1) + Y\right)^2 \lVert X\rVert_1^2\right) < \infty.
    \end{align}
    (\ref{Eqs: Poisson condition}) is satisfied when the covariate vector $X$ is Gaussian or sub-Gaussian distributed and the response $Y$ has a finite fourth moment.
\end{enumerate}
\end{proof}

\subsubsection{Gamma Regression}
\begin{proof} $ $
    \begin{enumerate}
    \item[A1]: $\ell_k(x, y; \cdot): \mathbb{T}_k \to \mathbb{R}$ is continuous for each $(x, y) \in \mathbb{X}$ as it is differentiable with respect to $\beta$. Fix $\beta \in \mathbb{T}_k$, $\ell_k(\cdot; \beta): \mathbb{X} \to \mathbb{R}$ is given by
    \begin{align*}
        \ell_k(x, y; \beta) &= \Tilde{f}(x, y; \beta)\Tilde{g}(x, y; \beta)  + \Tilde{h}_1(x, y; \beta) + \Tilde{h}_2(x, y; \beta)\\
        &= f(y; \beta)g(x; \beta)  + h_1(x; \beta) + h_2(y; \beta),
    \end{align*}
    where $\Tilde{f}(x, y; \beta) = f(y; \beta) = y$, 
    \begin{align*}
        \Tilde{g}(x, y; \beta) = g(x; \beta) &= \nu\exp(-\beta^\top x),\\
        \Tilde{h}_1(x, y; \beta) = h_1(x; \beta) &= \nu\beta^\top x + \log(\Gamma(\nu)) - \nu\log(\nu),\\
        \Tilde{h}_2(x, y; \beta) = h_2(y; \beta) &= - (\nu-1)\log(y).
    \end{align*}
    Both $g$ and $h_1$ are continuous functions, so they are measurable with respect to the Borel $\sigma$-algebra on $\mathbb{R}^d$. Both $f$ and $h_2$ are measurable with respect to the Borel $\sigma$-algebra on $\mathbb{R}_+$. So  $\Tilde{f}$, $\Tilde{g}$, $\Tilde{h}_1$ and $\Tilde{h}_2$ are measurable on the product space $\mathbb{X}$. It follows that $\ell_k(\cdot; \beta)$ is measurable on $\mathbb{X}$. 
    \item[A2]: $\mathbb{T}_k$ is compact. Let $\tau_k = 0 \in \mathbb{R}^d$, then $\tau_k \in \mathbb{T}_k$ for all $k \in \left[m\right]$. This condition is satisfied if $Y$ has a finite second moment,
    \begin{align*}
        \mathbb{E}\left(\ell_k(X, Y; \tau_k)^2\right) &= \mathbb{E}\left(\left(-\log (\Gamma(\nu)) + \nu\log(\nu) + (\nu - 1) \log (Y) - Y\nu\right)^2\right) \\
        &< \infty.
    \end{align*}
    \item[A3]: Fix $(x, y)\in \mathbb{X}$, $k \in \left[m\right]$ and $\beta, \beta'\in \mathbb{T}_k$. By the multivariate Mean Value Theorem,
    \[|\ell_k(x, y; \beta) - \ell_k(x, y; \beta')| \leq \sup_{\beta^*} \left\lVert \left. \frac{\partial \ell_k}{\partial \beta}\right|_{\beta = \beta^*} \right \rVert_2 \lVert \beta - \beta'\rVert_2, \]
    where $\beta^*$ is any point on the line segment joining $\beta$ and $\beta'$. We write $\beta^* = \lambda\beta + (1 - \lambda)\beta'$, for $\lambda\in\left[0, 1\right]$. 
    \begin{align*}
        \left.\frac{\partial \ell_k}{\partial \beta}\right|_{\beta=\beta} = \left(1 - y\exp(-\beta^\top x)\right)\nu x.
    \end{align*}
    Using triangle inequality,
    \begin{align*}
        \sup_{\beta^*} \left\lVert \left. \frac{\partial \ell_k}{\partial \beta}\right|_{\beta = \beta^*} \right \rVert_2 &= \sup_{\lambda\in\left[0, 1\right]} \left\lVert \left(1 - y\exp\left(-(\lambda\beta + (1 - \lambda)\beta')^\top x\right)\right)\nu x\right\rVert_2\\
        &= \sup_{\lambda\in\left[0, 1\right]} \nu\left\lvert1 - y\exp\left(-(\lambda\beta + (1 - \lambda)\beta')^\top x\right)\right\rvert \lVert x\rVert_2 \\
        &\leq \nu\left(1 + y\exp\left(\lvert\beta^\top x\rvert + \lvert\beta'^\top x\rvert\right)\right) \lVert x\rVert_1 \\
        &\leq \nu\bigr(1 + y\exp(\lVert\beta\rVert_1\lVert x\rVert_1 + \lVert\beta'\rVert_1\lVert x\rVert_1)\bigr) \lVert x\rVert_1\\
        &\leq \nu\bigr(1 + y\exp(2C_k\lVert x\rVert_1)\bigr) \lVert x\rVert_1.
    \end{align*}
    Let $\mathcal{G}_k(x, y) = \nu\bigr(1 + y\exp(2C_k\lVert x\rVert_1)\bigr) \lVert x\rVert_1$, we require the pair of covariates and response $(X, Y)$ to satisfy
    \begin{equation*}\label{Eqs: Gamma condition}
    \mathbb{E}\left(\mathcal{G}_k(X, Y)^2\right) = \mathbb{E} \left(\nu^2 \bigr(1 + Y\exp(2C_k\lVert X\rVert_1)\bigr)^2 \lVert X\rVert_1^2 \right)< \infty.
    \end{equation*}
    Similar to the Poisson regression, this is satisfied when the covariate vector $X$ is Gaussian or sub-Gaussian distributed and the response $Y$ has a finite fourth moment.
\end{enumerate}
\end{proof}

\subsection{Proof of Proposition \ref{Prop: BIC-like consistency}}

\begin{proof}
    By Theorem 2 of \cite{nguyen_panic:_2023}, it suffices to show conditions C1--C5. We start with C1. Using the functional form of $\ell_k$ from Table \ref{Table: Regression}, we have
    \begin{align*}
        r_k(\beta) &= \mathbb{E}(\ell_k(\beta; X, Y))\\
        &= \mathbb{E}\left(Y - \beta^\top X\right)^2\\
        &= \mathbb{E}(Y^\top Y) - 2\beta^\top \mathbb{E}(X Y) + \text{tr}(\beta\beta^\top\Sigma) + \mu^\top \beta\beta^\top\mu,
    \end{align*}
    where $\mu$ is the mean and $\Sigma$ be the covariance matrix of $X$, respectively. Differentiating $r_k$ twice, we get
    \begin{align*}
        \frac{\partial^2 r_k}{\partial \beta\partial\beta^\top}(\beta) &= 2 \Sigma + 2\mu\mu^\top.
    \end{align*}
    By our assumption that $\Sigma$ is a positive definite matrix, we see that the Hessian of $r_k$ is also positive definite. Hence, $r_k$ is strictly convex. As $r_k$ is strictly convex, it has, on a convex set, at most one minimizer, which we assume exists. To show Lipschitz continuity, first, we note that our assumptions on $X$ and $Y$ satisfy the conditions in Proposition \ref{Prop: Sufficient conditions}. This gives us the Lipschitz continuity of $\ell_k(\cdot; x, y)$, for all $(x, y)\in\mathbb{R}^d\times\mathbb{R}$. Then, for $\beta, \beta' \in \mathbb{T}_k$, 
    \begin{align*}
        \left|r_k(\beta) - r_k(\beta')\right| &\leq  \int_{\mathbb{R}^d \times \mathbb{R}} \left| \ell_k(\beta; x, y) - \ell_k(\beta'; x, y)\right| \, \Pi(dx, dy)\\
        &\leq \int_{\mathbb{R}^d \times \mathbb{R}} \mathcal{G}(x, y) \lVert \beta - \beta'\rVert_1 \, \Pi(dx, dy)\\
        &= c\lVert \beta - \beta'\rVert_1,
    \end{align*}
    for some $c < \infty$. This shows $\text{C1}$. Next, we note that, in our current setting, $\text{C2}$ is implied by condition $\text{A3}$. To show $\text{C5}$, let's first assume $\mathbb{T}_{k^*} \subset \mathbb{T}_{k}$. Since $r_k$ and $r_{k^*}$ are the same strictly convex function, $\beta_k^*$ and $\beta_{k^*}^*$ must be identical. The same argument can be applied to the case where $\mathbb{T}_{k^*} \supset \mathbb{T}_{k}$. Hence, condition $\text{C5}$ is satisfied. In the case of Lasso regression, condition $\text{C3}$ follows from the fact that $\mathbb{T}_k$ is polyhedral, for all $k\in\mathbb{R}_+$. In the case of the Ridge regression, this condition is justified using the Mangasarian--Fromovitz constraint qualification which holds for $\mathbb{T}_k$, for all $k\in\mathbb{R}_+$. Finally, in the case of Elastic-net regression, note the set $\mathbb{T}_k$ can be constructed by intersecting sets that satisfy the Mangasarian--Fromovitz constraint qualification, which leads to C3 by Proposition 3.90 from \cite{bonnans_perturbation_2000}. Condition $\text{C4}$ is satisfied if the Hessian of $r_k$ is positive definite, see \citet[p.25]{nguyen_panic:_2023}.  Hence, $\text{C4}$ is satisfied. 
\end{proof}

\textit{Remark}: The condition that the covariance matrix $\Sigma$ of $X$ is positive definite can be elaborated upon further. Suppose $\Sigma$ takes the form
\[\Sigma = \begin{bmatrix}
    0 & 0\\
    0 & \Omega
\end{bmatrix},\]
which is a representation of the covariance matrix of $\Bar{X} = (1, X^\top)$, the covariate vector augmented with a bias term. Suppose that $\Omega$ is a positive definite matrix in $\mathbb{R}^{d\times d}$, one can show that $\Sigma$ is positive definite. Indeed, let $M$ be the rank one matrix defined by
\begin{align*}
    M &= \mu\mu^\top\\
    &= \begin{bmatrix}
        a\\
        b
    \end{bmatrix}\begin{bmatrix}
        a\\
        b
    \end{bmatrix}^\top\\
    &= \begin{bmatrix}
        a^2 & ab^\top\\
        ab & bb^\top
    \end{bmatrix},
\end{align*}
where $a \in \mathbb{R}$ and $b \in \mathbb{R}^d$. Let $z = (x, y)^\top \neq 0$, where $x\in \mathbb{R}$ and $y \in \mathbb{R}^d$. Observe that
\begin{align*}
    z^\top (\Sigma + M)z &= \begin{bmatrix}
        x& y^\top
    \end{bmatrix} \left(
    \begin{bmatrix}
        0 & 0\\
        0 & \Omega
    \end{bmatrix} + 
    \begin{bmatrix}
        a^2 & ab^\top\\
        ab & bb^\top
    \end{bmatrix}
    \right)\begin{bmatrix}
        x\\
        y
    \end{bmatrix}\\
    &= \begin{bmatrix}
        x& y^\top
    \end{bmatrix}
    \begin{bmatrix}
        a^2 & ab^\top\\
        ab & \Omega + bb^\top
    \end{bmatrix}
    \begin{bmatrix}
        x\\
        y
    \end{bmatrix}\\
    &= \begin{bmatrix}
        a^2x + ay^\top b & axb^\top + y^\top\Omega + y^\top bb^\top
    \end{bmatrix}
    \begin{bmatrix}
        x\\
        y
    \end{bmatrix}\\
    &= a^2x^2 + 2axy^\top b + y^\top\Omega y + y^\top bb^\top y.
\end{align*}
For any value of $x$ and $y$, we have
\[a^2x^2 + 2axy^\top b + y^\top\Omega y + y^\top bb^\top y \geq 0.\]
If $y\neq 0$, the equality is strict as $\Omega$ is positive definite. So it suffices to check that the quadratic form is positive for $x\neq 0$ and $y = 0$. Note that,
\[a^2x^2 + 2axy^\top b + y^\top\Omega y + y^\top bb^\top y = a^2 x^2 > 0\]
for all $a \neq 0$. In particular, it is true when $a = 1$.

\subsection{Proof of Lemma \ref{Lemma: continuity}}
\begin{proof}
A finite-valued convex function is continuous on $\mathbb{R}^d$ \cite[Cor. 10.1.1]{rockafellar_convex_1997}, so it suffices to show the $\argmin$ function is continuous. The continuity of the $\argmin$ function results from the Berge Maximum Theorem \cite[Thm. 17.31]{aliprantis_infinite_2006}. Let $\phi: \mathbb{R}_{+} \twoheadrightarrow \mathbb{R}^d$ be the constant correspondence 
\[\phi(\lambda) = \{\beta\in \mathbb{R}^d: g( \beta) \leq g(\hat{\beta})\},\]
where $\hat{\beta}$ is the unique minimizer of $f$ on $\mathbb{R}^d$. Since $\phi$ is a constant correspondence, it is continuous. Moreover, the set $\phi(\lambda) \subset \mathbb{R}^d$ is compact for every $\lambda > 0$. To verify compactness, it suffices to show that $\phi(\lambda)$ is bounded. First, if $g(\hat{\beta}) \leq c$, then $\phi(\lambda) \subseteq L_c$ and it is necessarily bounded. So let $c < g(\hat{\beta})$, note that this implies $L_c \subseteq \phi(\lambda)$. Let's suppose for the sake of contradiction that $\phi(\lambda)$ is unbounded. Then, by Theorem 8.4 of \cite{rockafellar_convex_1997}, there exists some non-zero recession direction $y \in \mathbb{R}^d$ in the recession cone of $\phi(\lambda)$ such that, for all $\beta \in \phi(\lambda)$ and $t \geq 0$, $\beta + ty \in \phi(\lambda)$. This implies $g(\beta + ty)$ must be non-increasing in $t$, for all $\beta \in \phi(\lambda)$. So if we choose a $\beta$ that is also in $L_c$, then it holds that $\beta + ty \in L_c$ for all $t \geq 0$. This contradicts the boundedness of $L_c$, so it must be true that $\phi(\lambda)$ is bounded.  

Now, let $L: \text{Gr}_\phi \to \mathbb{R}$ be defined as 
\[L(\lambda, \beta) = f(\beta) + \lambda g(\beta).\]
Note that $L$ is continuous and strictly convex in $\beta$. Define the value function $m: \mathbb{R}_+ \to \mathbb{R}$ as
\begin{align*}
    m(\lambda) &= \min_{\beta \in \phi(\lambda)} L(\lambda, \beta).
\end{align*}
Fix any $\lambda > 0$, by Lemma \ref{Lemma: monotonicity}, the unique optimal solution $\beta_\lambda$ of $L(\lambda, \beta)$ on $\mathbb{R}^d$ satisfies $g(\beta_\lambda) \leq g(\hat{\beta})$. Hence, $\beta_\lambda \in \phi(\lambda)$ and the argmin correspondence $\mu$ of the value function $m$ is single-valued. By the Berge Maximum Theorem, $\mu$ is an upper hemicontinuous correspondence. Then, by Lemma. 17.6 of \cite{aliprantis_infinite_2006}, $\mu$ is continuous as a function.
\end{proof}

\subsection{Proof of Lemma \ref{Lemma: monotonicity}}
We use the same proof technique as \cite{oneto_tikhonov_2016}, but for the general case of $f$ and $g$.
\begin{proof} $ $
    Define $K^{\lambda_1}$ as the minimum value of the following problem 
    \begin{align*}\label{Eqs: Constrained problem}
    \begin{split}
        \min \quad & \, f(\beta) + \lambda_1 g(\beta)\\
        \text{s.t.}\quad & \beta \in \mathbb{R}^d, 
    \end{split}
    \end{align*}
    so that 
    \[K^{\lambda_1} = f(\beta_{\lambda_1}) + \lambda_1 g(\beta_{\lambda_1}).\]
    Similarly, define 
    \begin{align*}\label{Eqs: Constrained problem}
    \begin{split}
        K^{\lambda_2} =  f(\beta_{\lambda_2}) + \lambda_2 g(\beta_{\lambda_2}).
    \end{split}
    \end{align*}
    Let us show the desired result by eliminating other outcomes.
    \begin{enumerate}
        \item $g(\beta_{\lambda_1}) < g(\beta_{\lambda_2})$: Suppose $f(\beta_{\lambda_1}) \leq  f(\beta_{\lambda_2})$, this implies 
        \[f(\beta_{\lambda_1}) + \lambda_2 g(\beta_{\lambda_1}) < f(\beta_{\lambda_2}) + \lambda_2 g(\beta_{\lambda_2}) = K^{\lambda_2}, \]
        which contradicts the definition of $K^{\lambda_2}$. Now, suppose $f(\beta_{\lambda_1}) >  f(\beta_{\lambda_2})$. Using the definition of $K^{\lambda_1}$, we have
        \begin{equation}\label{Eqs: K1 inequality}
            K^{\lambda_1} = f(\beta_{\lambda_1}) + \lambda_1 g(\beta_{\lambda_1}) \leq f(\beta_{\lambda_2}) + \lambda_1 g(\beta_{\lambda_2}). 
        \end{equation}
        The above inequality implies
        \[\lambda_1 \geq \frac{f(\beta_{\lambda_2}) - f(\beta_{\lambda_1})}{g(\beta_{\lambda_1}) - g(\beta_{\lambda_2})}. \]
        However, following the definition of $K^{\lambda_2}$, we also know that 
        \begin{equation}\label{Eqs: K2 inequality}
            K^{\lambda_2} = f(\beta_{\lambda_2}) + \lambda_2 g(\beta_{\lambda_2}) \leq f(\beta_{\lambda_1}) + \lambda_2 g(\beta_{\lambda_1}), 
        \end{equation}
        which implies
        \begin{align*}
        \lambda_2 &\leq \frac{f(\beta_{\lambda_1}) - f(\beta_{\lambda_2})}{g(\beta_{\lambda_2}) - g(\beta_{\lambda_1})}\\
        &\leq \lambda_1.
        \end{align*}
        This contradicts the assumption that $\lambda_1 < \lambda_2$. Hence, $g(\beta_{\lambda_1}) < g(\beta_{\lambda_2})$ is impossible.
        \item $g(\beta_{\lambda_1}) > g(\beta_{\lambda_2})$: A similar argument to that used for the case $g(\beta_{\lambda_1}) < g(\beta_{\lambda_2})$ shows that $f(\beta_{\lambda_1})\geq f(\beta_{\lambda_2})$ is impossible. Let us show $f(\beta_{\lambda_1})< f(\beta_{\lambda_2})$ is plausible. Using (\ref{Eqs: K1 inequality}), here we have
        \[\lambda_1 \leq \frac{f(\beta_{\lambda_2}) - f(\beta_{\lambda_1})}{g(\beta_{\lambda_1}) - g(\beta_{\lambda_2})}.\]
        Similarly, (\ref{Eqs: K2 inequality}) implies that
        \[ \frac{f(\beta_{\lambda_1}) - f(\beta_{\lambda_2})}{g(\beta_{\lambda_2}) - g(\beta_{\lambda_1})}\leq \lambda_2.\]
        This observation is consistent with our original assumption that  $\lambda_1\leq\lambda_2$.
        \item $g(\beta_{\lambda_1}) = g(\beta_{\lambda_2})$: The case $f(\beta_{\lambda_1}) \neq f(\beta_{\lambda_2})$ is impossible, as it would lead to a contradiction with the definitions of $K^{\lambda_1}$ or $K^{\lambda_2}$. As for the case $f(\beta_{\lambda_1}) = f(\beta_{\lambda_2})$, first let us suppose that $\beta_{\lambda_1} \neq \beta_{\lambda_2}$. This is also impossible, as the number of optimal solutions is at most one. The remaining outcome $\beta_{\lambda_1} = \beta_{\lambda_2}$ is plausible. For example, let $f(\beta) = \lVert \beta\rVert_2^2$, $g(\beta) = \lVert \beta\rVert_1$. Then, for any $\lambda_1, \lambda_2 > 0$, 
        \[0 = \argmin_{\beta\in\mathbb{R}^d} f(\beta) + \lambda_1 g(\beta) = \argmin_{\beta\in\mathbb{R}^d} f(\beta) + \lambda_2 g(\beta).\]
    \end{enumerate}
\end{proof}

\bibliographystyle{plainnat}
\bibliography{citations}

\end{document}